\documentclass{aastex62}

\graphicspath{{./}{figures/}}
\usepackage{longtable}
\usepackage{url}
\usepackage{color}
\received{}
\revised{}
\accepted{}
\submitjournal{ApJ}

\shorttitle{3D evolution of CMEs}
\shortauthors{Majumdar et al.}

\begin{document}

\title{Connecting 3D evolution of Coronal Mass Ejections to their Source Regions}

\correspondingauthor{Dipankar Banerjee}
\email{dipu@iiap.res.in, dipu@aries.res.in}

\author[0000-0002-6553-3807]{Satabdwa Majumdar}
\affil{Indian Institute of Astrophysics,2nd Block, Koramangala, Bangalore, 560034, India}

\author[0000-0002-6954-2276]{Vaibhav Pant}
\affiliation{Department of Mathematics, Centre for mathematical Plasma Astrophysics, KU Leuven, Celestijnenlaan 200B, 3001 Leuven, Belgium}


\author[0000-0001-8504-2725]{Ritesh Patel}
\affiliation{Indian Institute of Astrophysics,2nd Block, Koramangala, Bangalore, 560034, India}
\affiliation{Aryabhatta Research Institute of Observational Sciences, Beluwakhan, 263001, Uttarakhand }

\author[0000-0003-4653-6823]{Dipankar Banerjee}
\affiliation{Indian Institute of Astrophysics,2nd Block, Koramangala, Bangalore, 560034, India}
\affiliation{Aryabhatta Research Institute of Observational Sciences, Beluwakhan, 263001, Uttarakhand }
\affil{Center of Excellence in Space Science, IISER Kolkata, Kolkata 741246, India}





\begin{abstract}

Since Coronal Mass Ejections (CMEs) are the major drivers of space weather, it is crucial to study their evolution starting from the inner corona. 
In this work we use Graduated Cylindrical Shell (GCS) model to study the 3D evolution of 59 CMEs in the inner ($<$ 3R$_{\odot}$) and outer ($>$ 3R$_{\odot}$) corona using observations from COR-1 and COR-2 on-board Solar TErrestrial RElations Observatory (STEREO) spacecraft. We identify the source regions of these CMEs and classify them as CMEs associated with Active Regions (ARs), Active Prominences (APs), and Prominence Eruptions (PEs). We find 27 $\%$ of CMEs show true expansion and 31 $\%$ show true deflections as they propagate outwards. Using 3D kinematic profiles of CMEs, we connect the evolution of true acceleration with the evolution of true width in the inner and outer corona. Thereby providing the observational evidence for the influence of the Lorentz force on the kinematics to lie in the height range of $2.5-3$ R$_{\odot}$. We find a broad range in the distribution of peak 3D speeds and accelerations ranging from 396 to 2465 km~s$^{-1}$ and 176 to 10922 m~s$^{-2}$ respectively with a long tail towards high values coming mainly from CMEs originating from ARs or APs. Further, we find the magnitude of true acceleration is be inversely correlated to its duration with a power law index of -1.19. We believe that these results will provide important inputs for the planning of upcoming space missions which will observe the inner corona and the models that study CME initiation and propagation.

\end{abstract}

\keywords{Sun: Corona - Sun: Coronal Mass Ejections (CMEs)}


\section{Introduction} 

Coronal Mass Ejections (CMEs) are large structures of plasma with embedded magnetic fields that are ejected from the solar atmosphere into the heliosphere \citep{article}. They appear as discrete, bright, white light feature propagating outwards in the coronagraph field of view (FOV) \citep{1984JGR....89.2639H}. CMEs apart from being such dynamics events, they are also the major drivers of space weather as they are capable of producing events like transient interplanetary disturbances, shock waves \citep{1993JGR....9818937G}, Thus, it is importaant to understand their kinematics. CMEs have been studied for several decades, with their observations dating back to the 1970s \citep{1971PASAu...2...57H}.
\citet{Dere1997SPD} made first observation of the CME from its initiation to propagation in the outer corona while combining data from {\textit{Extreme ultra-violet Imaging telescope}} (EIT) \citep{SOHOEIT} and {\it Large Angle Spectroscopic COronagraph} (LASCO) \citep{Brueckner95} on board {  \textit{Solar and Heliospheric Observatory}} (SOHO).
\citet{Cyr1999JGR} investigated the kinematics of 246 CMEs observed between 1980 and 1989 combining the observations from the inner ( $<$~2R$_{\odot}$) to outer ($>$~2R$_{\odot}$) corona using MK-3 K-coronameter (FOV: 1.12 - 2.44 R$_{\odot}$) at  Mauna Loa Solar Observatory (MLSO) and Solar Maximum Mission (SMM) \citep[1.5 - 6~R$_{\odot}$;][]{SMM1980}. It was proposed that a driving force is responsible for the continuous acceleration and expansion of CMEs up to the FOV of the SMM. Later, it was reported that CMEs pose three-phase kinematic profile with an initial slow gradual rise phase, an impulsive acceleration phase below 2 R$_{\odot}$, and the final phase with constant or decreasing speed \citep{Zhang2001, Zhang_2004,2011ApJ...738..191B}. Beyond this, a CME propagates in the interplanetary medium with almost constant speed or show gradual acceleration/deceleration due to the interaction with the ambient solar wind \citep{nat2000}. Thus, an anti-correlation between the average speed and acceleration, interpreted as an effect of aerodynamic drag,  was found in height between 2-30 R$_\odot$ for over 5000 CMEs \citep{vrsnak2004A&A}. Furthermore, it was found that during the CME propagation, greater acceleration magnitudes (A) have small duration (T) and vice versa following A $\sim$ T$^{-1}$  \citep{Zhang_2006}. 

The study of the observable properties of CMEs using a large sample from SOHO provided an understanding on the nature of CMEs and variation with the solar cycle \citep{StCyr2000JGRS, Yashiro04, Gopalswamy2006JApAG,2007SoPh..241...85V, Gopalswamy2009EM&PG}.

CMEs are observed to be associated with flares and eruptive prominences \citep{article}. The kinematics of CMEs associated with these two classes are investigated in earlier studies \citep{Sheeley1999JGR,Moon2002ApJM, Chen2006A&AC, Bein2012ApJB}. \citet{Vrsnak2005A&AV} studied a sample of 545 CMEs and reported a contrasting result demonstrating that both the CME classes have similar kinematics characteristics. It has been observed that the CME kinematics is governed by the interplay of three forces, namely the Lorentz force, the gravitational force and the viscous drag force which is manifested in a wide range in their kinematics with their speeds ranging from a few tens to thousands km~s$^{-1}$ and their acceleration ranging from a few tens to even a few 10$^4$ m~s$^{-2}$ \citep{Wood1999ApJW, Zhang2001, article,2007SoPh..241...85V}. Also, CMEs attain enormous acceleration in the inner corona where a strong magnetic field generates large Lorentz force \citep{2007SoPh..241...85V,2011ApJ...738..191B}. Recently, \citet{2020arXiv200403790C} studied the slow rise and main acceleration phase of 12 CMEs using the data from {\textit{Atmospheric Imaging Assembly}} (AIA) on-board {  \textit{Solar Dynamics Observatory}} (SDO) \citep{aia} and EUVI on-board STEREO and suggested that the main acceleration phase is qualitatively different from the slow rise phase and that different physical mechanisms govern these two phases. 
Hence to understand this initial impulsive acceleration and the forces involved in propelling the CME, it is important to understand the kinematics of CMEs in the inner corona ($<$2~R$_{\odot}$).

Apart from radial propagation, CMEs also exhibit lateral expansion leading to a increase in the width from the inner to outer corona (\citet{Kay_2015}, and references therein). In earlier studies of CMEs the variation of radial and expansion speeds was found and shown that it was dependent on the width of the CMEs \citep{MacQueen1983SoPhM, 2009CEAB...33..115G}.

The lateral expansion of CMEs is due to the injected energy from the magnetic field of the source region.

The Lorentz force from this injected magnetic energy gives the necessary thrust that translates the CME and leads to its expansion \citep{Subramanian_2014, Suryanarayana2019JASTP}. 

Recently \citet{2020A&A...635A.100C} studied 12 CMEs using the data from the SDO/AIA, SOHO/LASCO, STEREO/SECCHI coronagraphs and inferred from 3D reconstruction that the initial true expansion of CMEs below 3 R$_{\odot}$ is asymmetric and non-self-similar.

It should be noted that in the studies based on LASCO-C1, EUVI or AIA to probe inner corona, the images observed in emission lines were used to track CMEs. It is still debatable if the same features are observed in emission lines and white-light.
Moreover, a major limitation of all these studies is that the properties of CMEs, width, speed, acceleration, propagation direction, are measured using observations from single view-point. As a result these values measured in the plane of sky suffer from projection effects \citep{Cremades2004A&A, Burkepile2004JGRAB, Temmer2016AN, Balmaceda2018ApJ}. 
To overcome such limitation, \citet{Sarkar2019ApJ} (and references therein) exploited the coronal cavity observed in EUV and white-light images and found that the spatial relationship is preserved while tracking the feature in 3D in these pass-bands. 
Further to minimise the projection effects, \citet{Mierla2008SoPh} derived the 3D-kinematics of CMEs using STEREO/COR-1, COR-2 images from two vantage points by a triangulation method and compared it with other methods \citep{Mierla2009SoPh, Mierla2010AnGeo, Mierla2011JASTP}. Using polarimetric capability of COR-1 A and B coronagraphs, \citet{Moran2010ApJ} determined the 3D orientation of a CME.
Another method based on forward modelling to fit the CME flux rope on multi vantage point images was developed assuming the self-similar expansion of CMEs \citep{Thernisien2006ApJ, 2009SoPh..256..111T, Thernisien2011ApJST}. An automated method based on triangulation to identify and track 3D structure of CMEs is recently developed by \citet{Hutton2017A&A}.

One of the first 3D kinematical study of CMEs combining the data from STEREO/EUVI and coronagraphs was done by \citet{Joshi2011ApJ}.
 \citet{2012SoPh..281..167B} derived and catalogued 3D properties of 51 CMEs from January 2007 to December 2010 using the data from STEREO/COR-2.
\citet{Sachdeva_2015} used a sample of 8 CMEs observed by SOHO/LASCO C2, STEREO/COR2 and STEREO/HI to study their propagation in 3D and the impact of aerodynamic drag, and further extended their study in \citet{2017SoPh..292..118S} to 38 events to understand the relative importance of Lorentz force and drag force, but their height measurements left out the information in the inner corona.
The multi-viewpoint analysis of a CME observed on March 28, 2013 yielded a linear increase in angular width of CME up to 5 R$_\odot$ \citep{Cabello2016SoPh}.

A sample of 460 CMEs observed by twin STEREO spacecrafts was used to compare the single-viewpoint and 3D kinematics (via triangulation) of CMEs in COR-2 FOV \citep{Balmaceda2018ApJ}.  
It is worth noting that although these studies provide 3D kinematic information, most of the analysis (\citet{2020A&A...635A.100C} and \citet{Cabello2016SoPh} being exceptions in this context) are carried out starting from the COR-2 FOV, and hence the essential information in the inner corona is left out.  Thus the major gap in our understanding of CME kinematics is to understand their 3D evolution in the inner corona.

During the early evolution phase, CMEs often show non-radial deflections in their trajectories \citep{nat,2011SoPh..271..111G,Lugaz_2012,2019arXiv190906410W}. It has been observed that CMEs can get deflected from their initial path  when they get ejected near a coronal hole \citep{nat, Kahler_2012}. 

 \citet{2013ApJ...775....5K} developed a tool, ForeCAT (Forecasting a CME's Altered Trajectory) for studying CME deflections due to the magnetic forces.
\citet{Kay_2015} (and references therein), reported on observed latitudinal and longitudinal deflections suffered by CMEs as they propagated outwards. A study of these deflections is important for a better understanding of their trajectories and also for better space weather forecasting \citep{Kay_2015}.

As it has been established that the Lorentz force is responsible for propelling and expanding the CME \citep{Subramanian_2014}, a connection between the evolution of the width of CMEs with their kinematic profiles will shed light on the imprint of Lorentz force on the evolution of CMEs.
In this work, to gain a better understanding of the kinematics of CMEs, we perform a statistical study of 3D evolution of 59 CMEs in the inner and outer corona occurring between May 2007 to April 2014 and connect them to their source region. To keep uniqueness in the physical feature we track the CME only in white light coronagraph data.
Since the use of triangulation method involves a smaller feature (point) reconstructed in 3D, we used the Graduated Cylindrical Shell (GCS) model developed by \citet{Thernisien2006ApJ, 2009SoPh..256..111T} which fits a flux rope to entire CME structure using a pair of observed images by the coronagraphs COR-1 and COR-2 on-board STEREO spacecrafts from two different points. In this work we track the 3D structure of CMEs and try to understand their true evolution during the propagation. The paper is organised as follow: in Section \ref{sec2} we outline the data sources that have been used followed by the working method. The results of our analysis are presented in Section \ref{sec3} followed by summary and conclusions in Section \ref{sec:summary}.

\section{Data and Method}\label{sec2}
\subsection{Data Source and Data Preparation}

We have used the Coordinated Data Analysis Workshop (CDAW) catalogue, which records the properties of CMEs detected manually in the SOHO/LASCO images \citep{Gopalswamy2009EM&PG} to select events.
The data from different passbands of Atmospheric Imaging Assembly (AIA) on-board Solar Dynamics Observatory (SDO) \citep{aia} and Extreme ultraviolet Imaging Telescope (EIT) \citep{SOHOEIT} on-board SOHO were used to identify the source regions of CMEs that were coming from the front side of the Sun.  For events happening at the back side of the Sun, data from EUVI on-board Sun Earth Connection Coronal and Heliospheric Investigation (SECCHI) package \citep{secchi} of the twin spacecraft STEREO-A and STEREO-B was used.
Since we aim to understand the 3D evolution of CMEs in the inner corona, high cadence data from multiple vantage points are needed. The COR-1 coronagraph on-board STEREO-A and STEREO-B offers high cadence (5 minutes) observation of the inner corona from two vantage points, which can be used to reconstruct the 3D structure of the CME. In this regard, the data from COR-1 with field of view 1.5 - 4 R$_{\odot}$ and COR-2 having field of view from 2.5 - 15 R$_{\odot}$  of the  SECCHI package on-board STEREO-A and STEREO-B was combined to study the kinematics of CMEs close to their initiation in the inner corona and their further evolution. Level 0.5 data of EUVI, COR-1 and COR-2 was reduced to level 1 using $secchi\_prep.pro$ in IDL. Next, base difference images were created by subtracting a pre-event image from successive images of the event, thus ensuring that we are tracking the same feature in all images. 

\subsection{Event selection and Source Region identification}
We select CMEs from the CDAW catalogue, occurring during the period from May 2007 to April 2014 which includes the period of minimum sunspots in solar cycle 23 to the period of maximum sunspots of cycle 24, and which are bright in STEREO data, so that GCS model can be fitted with reasonable accuracy. We selected CMEs which showed a distinct leading edge that can be clearly tracked in successive images of COR-1 and COR-2. We did not select CMEs which were marked as "very poor" in the CDAW catalogue because such CMEs have a very faint leading edge which is not clearly seen in the successive frames of COR-1 and COR-2 images, thus tracking them can be ambiguous. In this regard, we note that there are also other automated CME catalogues like SEEDs \citep{seeds}, CACTUS \citep{cactus,2016ApJ...833...80P}, ARTEMIS \citep{artemis}, CORIMP \citep{corimp_1,corimp_2},  and others. JHUAPL \citep{jhuapl} identifies the events visually, the measurements are however estimated by a semi-automatic algorithm. But Since, we choose bright events in our analysis so that GCS model can be fitted with reasonable accuracy, these events are likely to be captured by both manual and automated catalogs. It seems it does not matter which catalog we use to identify the CMEs. Having said that, we understand that the estimates of the kinematic properties (such as velocity, width, and accelerations) might be different in different catalogs due to different measurement techniques used.  But, our conclusions are not based on any catalogued parameters, and is completely based on the computed GCS parameters. Thus, the choice of catalogue will not affect our results. Only in Section \ref{compar} we show a comparison of projected and true speeds to show the importance of true speeds over projected speeds. For this we use CDAW speed because CDAW tracks the leading edge manually, and our work also involves manual fitting of the GCS model. Thus we believe that using automated catalogs instead of manual catalog will not significantly alter the results of this study. Motivated by the facts that CMEs show a wide range in their speeds \citep{article} and that the average solar wind speed naturally divides CMEs into slow and fast ones \citep{nat2000}, we segregate the selected CMEs into slow ($<400$ km~s$^{-1}$) and fast ($>400$ km~s$^{-1}$) based on the average solar wind speed, which is around 400 km~s$^{-1}$ \citep{2006LRSP....3....2S}.  After putting these constraints on our event selection, we then back project the CMEs on to the solar disk and with the help of the data from SOHO/EIT (195 \AA, 304 \AA), SDO/AIA (171 \AA, 193 \AA, 304 \AA), STEREO-A/EUVI-A (195 \AA, 304 \AA) and STEREO-B/EUVI-B (195 \AA, 304 \AA),  we identify their source regions. For associating a CME to its source region we follow a similar procedure as mentioned in \citet{Gilbert_2000}. We spatially associate a source region to a CME by requiring that the latitude of the source region to be around $\pm$ $30^{\circ}$ to that of the central position angle (PA) of the CME converted to equivalent apparent latitude ($lat_{PA}$) by the following relation:

\begin{equation}
    lat_{PA} = 90-PA \quad [0 \leq PA \leq 180] \qquad
    lat_{PA} = PA-270 \quad [180 < PA \leq 360].
\end{equation}

We  temporally associate them by requiring that the source region erupts or shows radially outward movement in the above latitude window within at least 30 minutes before the time of first appearance of the CME leading edge in the LASCO C2 field of view. We classify the identified source regions into three broad categories. Active Regions (ARs), which are regions of strong magnetic field showing bright emissions in EUV and X-rays, Prominence Eruptions (PEs), which are the quiescent prominences seen as dense hanging feature in the solar atmosphere and Active Prominences (APs) which are PEs with their one or more foot-points connected to ARs (also refer \citet{2001ApJ...561..372S}). After associating the CMEs to their source regions, we segregated the CMEs into 6 separate classes, fast CMEs from ARs, PEs, and APs and similarly for slow CMEs. 

\subsection{3D Reconstruction and Estimation of Geometrical Parameters}

\begin{figure*}[h]
\gridline{\fig{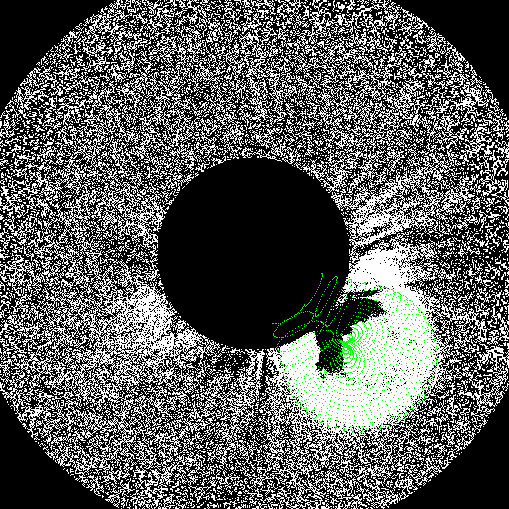}{0.4\textwidth}{(a)}
          \fig{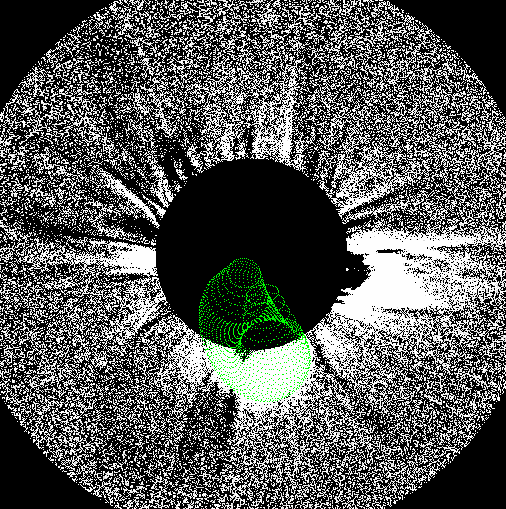}{0.4\textwidth}{(b)}
          }
\caption{The CME on $2012-11-09$ observed by COR-1 A and COR-1 B at 01:00:00 UT, with the GCS model generated flux-rope structure (green wireframe) overlaid on them. Note that the CME has a face on view in (a) and an edge on view in (b)} 
\label{fitting}
\end{figure*}

To study the 3D kinematics of these CMEs, we use the Graduated Cylindrical Shell (GCS) model developed by \citet{2009SoPh..256..111T}. This model assumes a flux-rope structure of CMEs. It provides 6 parameters: height of the leading front ($h$), longitude ($\phi$), latitude ($\theta$), tilt-angle ($\gamma$), aspect-ratio ($k$), and half-angle ($\alpha$), to fit the model to an event and reproduce the large scale flux-rope structure of the CME. We use this model by fitting a geometrical wireframe structure of the model simultaneously to a synchronized pair of SECCHI  (EUVI, COR-1 and COR-2) images. This geometrical structure resembles a hollow croissant with conical legs, and with a front section which is in the approximate shape of a torus. This cross-section ($a(r)$) increases with height (r) subject to the condition that the ratio of the former to the later ($k=a(r)/r$) is constant with time, where the $k$ is the aspect-ratio parameter. The fitting is done interactively, until a good visual match is obtained between the wireframe projection of the model and both the data views from COR-1 (A and B) or COR-2 (A and B). We start with a pair of image where the CME leading front is well developed. We set the half-angle to zero initially and adjust the latitude and longitude to set the location of the foot-point of the model. Then the height is adjusted until the wireframe covers the leading outer front of the CME. Next, we adjust the aspect-ratio to match the spatial extension of the CME. Finally, we change and adjust the half-angle and tilt angle until the best visual match is obtained. Figure \ref{fitting} shows an example of fitting the synthetic wireframe structure (in green) using GCS model to a pair of COR-1 images for the CME observed on 2012-11-09 at 01:00:00 UT.
To quantify how well we have fitted the model in reproducing the overall structure, the uncertainty in fitting is found by outlining a hand drawn contour on the stereoscopic pair of images (an automatic optimizer is further used to refine this fit) and then a comparison is done between the hand drawn contour and the binary contour of the projected wireframe structure of the model (for details refer \citet{2009SoPh..256..111T}). This whole process is repeated iteratively for a time sequence of images in the COR-1 and COR-2 field of view and the evolution of each of the parameters is recorded. In this context, \citet{2009SoPh..256..111T} reported that for some cases, it is not possible to determine the complete flux-rope orientation from observation from two vantage points (in particular for periods in which STEREO-A and STEREO-B are in/close to opposition.), use of a third vantage point will help in a better 3D reconstruction (also refer \citet{2018ApJ...863...57B}). In this regard, use of LASCO images can help in better constraining the parameters. But, since we want to study the evolution of CME in the inner corona, and LASCO C2 FOV starts from 2 R$_\odot$, we will be able to use LASCO only beyond 2 R$_\odot$, and below this height we have to use COR-1 only. Thus, to keep consistency throughout our analysis, we didn't use LASCO data in this work.

\subsection{Methods of the estimation of 3D kinematical properties}

After fitting the GCS model to a time sequence of images of COR-1 and COR-2, the fitted parameters are recorded. The height time data is fitted with a cubic smooth spline (red curve in top panel of Figure 2).  A R-based procedure ``smooth.spline.R" \citep{R} is used to perform cubic smooth spline fitting. In this method, the number of nodes are supplied by the user, and the fitting function then decides on the position of the nodes, thus dividing the entire height time data into those many segments. We choose on average, six to ten nodes to characterise the spline, depending on the number of data points present in  each event, thus if the number of data points were less, then the number of nodes was reduced to 6 or 7, such that the different phases of evolution can be captured. The speed and acceleration are calculated by the first and second order numerical differentiation of the height time data. The speed and acceleration profile (red curve in second and third panel) is similarly obtained by the numerical differentiation of the fitted height time smooth spline function. We also plot an inset with a zoom into the residual acceleration phase in the bottom corner of the panel displaying acceleration.\\
The grey band around the red curve in the top panel denotes the uncertainty in the fitted parameter which is estimated by computing the uncertainty in the fitting of the GCS model to the pair of observed images. The uncertainty in fitting is found by outlining a hand drawn contour on the stereoscopic pair of images (an automatic optimizer is further used to refine this fit) and then a comparison is done between the hand drawn contour and the binary contour of the projected wireframe structure of the fitted model (for details refer \citet{2009SoPh..256..111T} ). This procedure is repeated several times on several images and the average uncertainty was found to be 20\%. Thus the grey band denotes the 20\% uncertainty region.

 In section \ref{accel_dur_mag}, we study the acceleration magnitude and acceleration duration of the CMEs. The acceleration phase is determined from  the velocity profiles. We identify the starting time of acceleration (when the velocity starts changing) and the ending time of acceleration (when the velocity reaches the highest value and then becomes constant or decelerates from thereupon), from which we estimate the duration of the acceleration. The acceleration magnitude is then calculated (as reported by \citet{Zhang_2006}) as the increase in velocity divided by the acceleration duration. The uncertainty in the duration of the acceleration is due to the cadences of COR-1 and COR-2. The uncertainties in height measurements (as mentioned earlier) and the time interval (from cadence of COR 1 and COR 2) are used to compute the uncertainty in magnitude of the acceleration. Since the uncertainty of velocity change, and acceleration duration is inherited into the uncertainty of acceleration magnitude, the uncertainty is higher for impulsive events, and smaller for gradual events (also reported by \citet{Zhang_2006}). Apart from the image cadence, uncertainty in the acceleration phase can come from the inconsistencies of the CME leading edge as measured in different wavelengths (EUV and white light), and from the change in intensity over the FOV of a single instrument \citep{Temmer_2010}. As we have mentioned earlier, we have not used EUV data with white light data, to remove such inconsistencies in the measurement of the leading edge in different wavelengths.  We cannot rule out the acceleration below COR-1 FOV, and in such cases, the measurement of the duration of the acceleration phase would be an under estimate. However, including EUV data will be tricky since we do not know if the structures seen in EUV map to the same features observed in white-light. Thus we confined the analysis to white light only.

\section{Results}
\label{sec3}
The details of the GCS fitting is presented in Table~\ref{table2} with all fitting parameters at the time when the CME leading front is well developed in the COR-1 FOV. The third column specifies the identified source region of the CMEs. In case of events where both the view points gave an edge-on view of the CME, in those cases, we have an ice cream cone shape, which is reproduced by the model by setting $\alpha$ to zero, and in such cases we put a \lq$-$\rq~in the column for $\gamma$ in Table \ref{table2}. 
Different physical parameters of the model provide physical insights on their evolution. A change in the GCS latitude, or longitude is generally interpreted as true deflection, a change in $\gamma$ is generally interpreted as rotation as the CME propagates into the heliosphere, and a change in the half-angle can be interpreted as true 3D expansion of the CME \citep{2009SoPh..256..111T,2012SoPh..281..167B}\\
Once we record parameters of CME for a time series during the CME evolution through COR-1 and COR-2 FOV, we explore the physics conveyed by these parameters and their impacts on the evolution of CMEs through the inner and outer corona.

\begin{longrotatetable}
\begin{deluxetable*}{ccccccccccc}  
\setlength{\tabcolsep}{2pt}
\tablecaption{The GCS model parameters of all the CMEs are tabulated below. $h$ is the height of the leading front, $\theta, and \phi$ are the latitude and longitude of the foot-points of the CME, the tilt-angle ($\gamma$) is the angle between the axis of symmetry of the CME and the local parallel of latitude, aspect-ratio ($k$) is the ratio of the radius of cross-section of the leading front and the radial distance of the leading front from the center of the Sun and $\alpha$ is the half-angle between the legs of the best fit GCS model.  The Time is the time of observation, $V_{CDAW}$ is the average linear speed taken from the CDAW catalogue, and $V_{GCS}$ is the average 3-D speed calculated from the GCS model. \label{table2}}
\tablecolumns{9}
\tablenum{1}
\tablewidth{0pt}
\tablehead{
\colhead{Date} & 
\colhead{Time} & 
\colhead{{Source} } & \colhead{Height ($h$)} & \colhead{Longitude ($\phi$)} & \colhead{Latitude ($\theta$)} & \colhead{Tilt-angle ($\gamma$)} & \colhead{Aspect-ratio ($k$)} & \colhead{Half-angle ($\alpha$)} & \colhead{$V_{CDAW}$ } & \colhead{$V_{GCS}$}\\
\colhead{\footnotesize(YYYY-mm-dd)} & \colhead{\footnotesize(HH:MM:SS)} &
\colhead{Region} & \colhead{\footnotesize (R$_{\odot}$)} & \colhead{\footnotesize(Degree)} & \colhead{\footnotesize(Degree)} & \colhead{\footnotesize(Degree)} & \colhead{} & \colhead{\footnotesize(Degrees)}  & \colhead{\footnotesize(km s$^{-1}$)}  & \colhead{\footnotesize(km s$^{-1}$)} 
}
\startdata
2007-05-09 & 02:00:00 & AR & 3.36 & 69 & 3.9 & $-$ & 0.33 & 0 & 264 & 277 \\ 
2008-03-25 & 19:20:00 & AR & 3.36 & 188 & -15 & 69 & 0.17 & 12 & 1103 & 1074 \\
2008-03-26 & 10:52:22 & AR & 3.71 & 1 & -5 & 2 & 0.21 & 4 & 163 & 241\\
2008-04-05 & 16:15:00 & PE & 3.35 & 258 & 0 & -65 & 0.13 & 14 & 962 & 994\\
2008-04-09 & 10:45:00 & AP & 3.22 & 193 & -21 & 2 & 0.12 & 8 & 650 & 543\\
2008-10-17 & 05:25:00 & PE & 3.21 & 276 & -22 & -31 & 0.26 & 18 & 143 & 244\\
2010-02-13 & 00:30:00 & AR & 2.98 & 199 & 44 & 8 & 0.21 & 9 & 1005 & 325\\
2011-01-24 & 01:45:00 & PE & 3.39 & 0 & -24 & -36 & 0.22 & 11 & 258 & 301\\
2011-04-02 & 11:50:00 & AP & 3.29 & 119 & -43 & 61 & 0.19 & 30 & 238 & 300\\
2011-05-12 & 13:40:00 & AP & 3.37 & 0 & 18 & 40 & 0.13 & 11 & 274 & 390\\
2011-05-17 & 05:20:00 & PE & 3.36 & 239 & 13 & 90 & 0.13 & 12 & 220 & 231\\
2011-05-30 & 10:40:00 & PE & 3.25 & 273 & 61 & 90 & 0.19 & 27 & 299 & 254\\
2011-06-01 & 18:05:00 & AR & 3.29 & 38 & -5 & -43 & 0.38 & 5 & 198 & 378\\
2011-06-11 & 13:10:00 & PE & 3.31 & 252 & 30 & -58 & 0.15 & 18 & 269 & 258\\
2011-06-20 & 17:55:00 & PE & 3.21 & 28 & 16 & 7 & 0.28 & 19 & 329 & 470\\
2011-07-09 & 00:40:00 & AP & 3.29 & 278 & -17 & 57 & 0.17 & 21 & 209 & 692\\
2011-09-20 & 13:15:00 & AP & 3.22 & 234 & 24 & 68 & 0.17 & 31 & 337 & 427\\
2011-12-26 & 11:50:00 & AR & 3.24 & 186 & 24 & 62 & 0.20 & 25 & 448 & 782\\
2012-01-02 & 01:47:30 & AP & 3.34 & 265 & -19 & 90 & 0.28 & 39 & 531 & 540\\
2012-01-06 & 21:30:00 & AR & 3.33 & 279 & 51 & 24 & 0.24 & 20 & 443 & 407\\
2012-01-08 & 02:30:00 & AR & 3.27 & 240 & -31 & -45 & 0.35 & 37 & 557 & 530\\
2012-01-15 & 03:10:00 & AR & 3.20 & 224 & 39 & -52 & 0.26 & 36 & 407 & 526\\
2012-01-18 & 12:10:00 & AR & 3.21 & 226 & -30 & -21 & 0.28 & 22 & 267 & 367\\
2012-01-19 & 10:25:00 & AR & 3.21 & 278 & 46 & 51 & 0.13 & 22 & 270 & 363\\
2012-02-24 & 03:20:00 & PE & 3.11 & 107 & 25 & -64 & 0.20 & 38 & 800 & 473\\
2012-03-12 & 01:30:00 & PE & 3.15 & 181 & -53 & -3 & 0.09 & 29 & 638 & 357\\
2012-03-15 & 00:20:00 & AR & 3.15 & 335 & 12 & -38 & 0.19 & 30 & 750 & 354\\
2012-03-16 & 20:55:00 & PE & 3.20 & 263 & 43 & - & 0.20 & 0 & 862 & 515\\
2012-03-20 & 23:20:20 & AR & 3.00 & 36 & 53 & 22 & 0.18 & 6 & 253 & 324\\
2012-03-30 & 15:00:00 & PE & 3.08 & 318 & 29 & -71 & 0.27 & 30 & 584 & 506\\
2012-04-08 & 01:40:00 & PE & 3.25 & 13 & -27 & -17 & 0.40 & 7 & 624 & 565\\
2012-04-19 & 15:30:00 & AP & 3.29 & 104 & -30 & 90 & 0.16 & 11 & 540 & 495\\
2012-04-28 & 12:30:00 & PE & 3.14 & 26 & -38 & 3 & 0.37 & 17 & 260 & 213\\
2012-04-30 & 07:50:00 & AR & 3.21 & 51 & -28 & -27 & 0.18 & 32 & 992 & 935\\
2012-05-03 & 14:45:00 & AR & 3.16 & 200 & 14 & -75 & 0.58 & 6 & 584 & 554\\
2012-05-24 & 22:30:00 & AP & 3.21 & 358 & -5 & 45 & 0.14 & 15 & 563 & 405\\
2012-06-03 & 18:25:00 & AP & 3.42 & 197 & 26 & -73 & 0.25 & 24 & 605 & 710\\
2012-06-15 & 14:45:00 & AP & 3.07 & 275 & -9 & 57 & 0.22 & 59 & 262 & 502\\
2012-06-26 & 10:10:00 & AR & 3.43 & 103 & -22 & -10 & 0.24 & 14 & 283 & 367\\
2012-06-27 & 11:30:00 & PE & 3.29 & 32 & 53 & -3 & 0.33 & 9 & 511 & 441\\
2012-06-29 & 01:55:00 & PE & 3.15 & 231 & -12 & 77 & 0.11 & 18 & 207 & 437\\ 
2012-07-05 & 13:45:00 & AR & 3.28 & 229 & -42 & 10 & 0.25 & 27 & 741 & 623\\
2012-07-06 & 06:10:00 & AR & 3.07 & 287 & -11 & 73 & 0.23 & 29 & 258 & 570\\
2012-07-08 & 16:40:00 & AP & 3.57 & 225 & -38 & -53 & 0.24 & 7 & 1572 & 1492\\
2012-07-17 & 14:20:00 & AR & 3.16 & 69 & -32 & -21 & 0.21 & 20 & 958 & 504\\
2012-07-26 & 23:20:00 & PE & 3.15 & 161 & 58 & -18 & 0.53 & 34 & 542 & 421\\
2012-08-02 & 13:15:00 & PE & 3.01 & 274 & -26 & 45 & 0.41 & 25 & 563 & 551\\
2012-08-04 & 12:50:00 & AR & 3.12 & 93 & -20 & 50 & 0.20 & 38 & 187 & 318\\
2012-10-12 & 01:50:00 & AP & 3.14 & 166 & 56 & -58 & 0.30 & 10 & 275 & 228\\
2012-10-22 & 02:00:00 & AP & 3.17 & 341 & -17 & 90 & 0.12 & 14 & 304 & 300\\
2012-11-09 & 01:00:00 & PE & 3.13 & 182 & -38 & -49 & 0.27 & 17 & 771 & 664\\
2012-12-10 & 15:55:00 & AP & 3.07 & 114 & -13 & 65 & 0.22 & 32 & 305 & 424\\
2012-12-14 & 02:15:00 & PE & 3.14 & 232 & -17 & 56 & 0.18 & 30 & 304 & 440\\
2013-04-05 & 22:10:24 & AP & 3.22 & 234 & -5 & 90 & 0.19 & 14 & 207 & 306\\
2014-01-02 & 07:30:00 & PE & 3.00 & 235 & -48 & 79 & 0.17 & 23 & 168 & 230\\
2014-01-04 & 23:02:30 & AP & 3.13 & 237 & 1 & 89 & 0.23 & 34 & 567 & 559\\
2014-01-07 & 03:45:00 & AP & 2.99 & 259 & -31 & -90 & 0.40 & 61 & 688 & 619\\
2014-02-11 & 19:30:00 & AP & 3.00 & 105 & -19 & -32 & 0.32 & 26 & 613 & 658\\
2014-04-07 & 21:45:00 & PE & 3.03 & 107 & -28 & -75 & 0.22 & 18 & 160 & 512\\
\hline
\enddata
\end{deluxetable*}
\end{longrotatetable}

\subsection{Connecting Width to the 3D Kinematics} \label{sec6}

\begin{figure*}[!ht]
\gridline{\fig{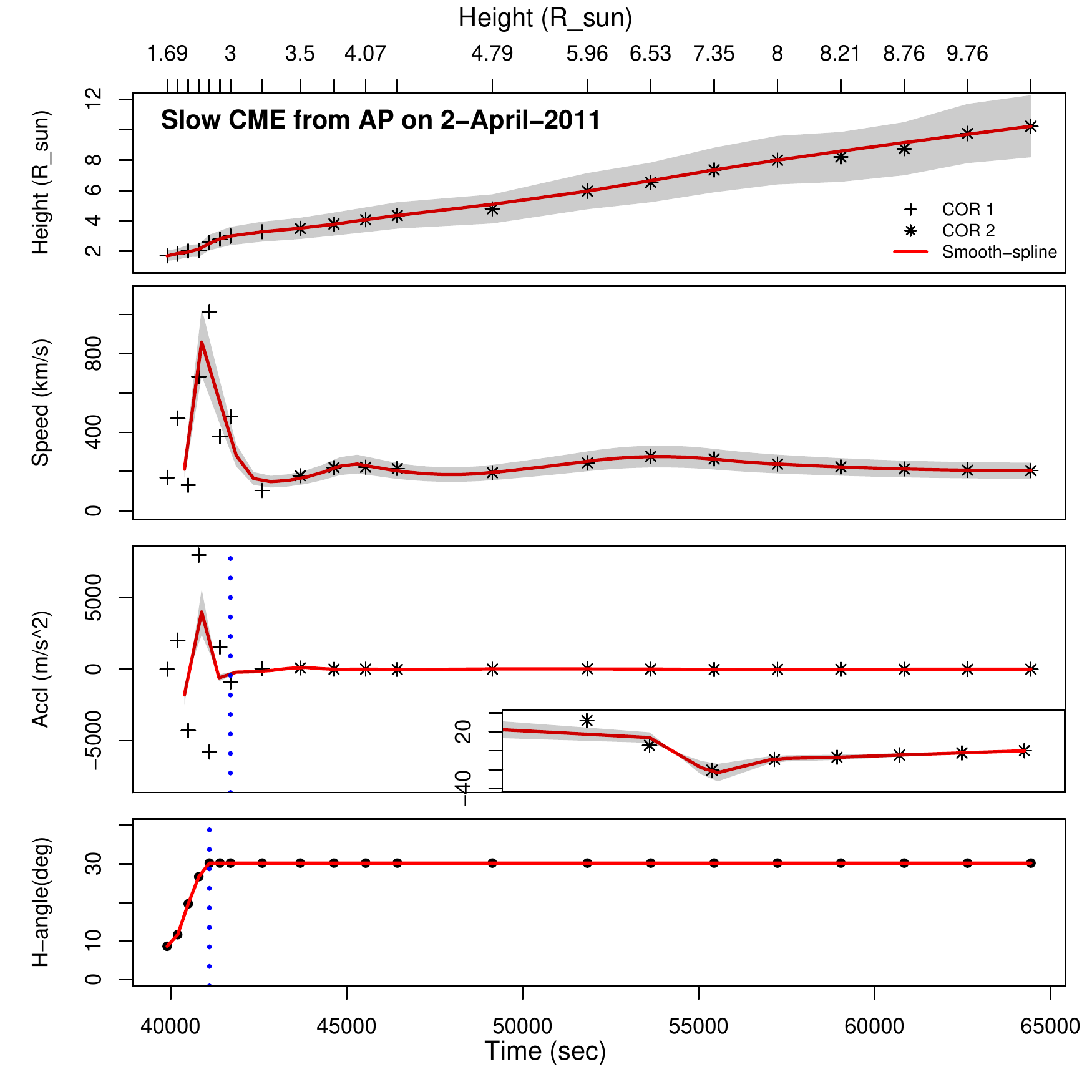}{0.52\textwidth}{(a)}
          \fig{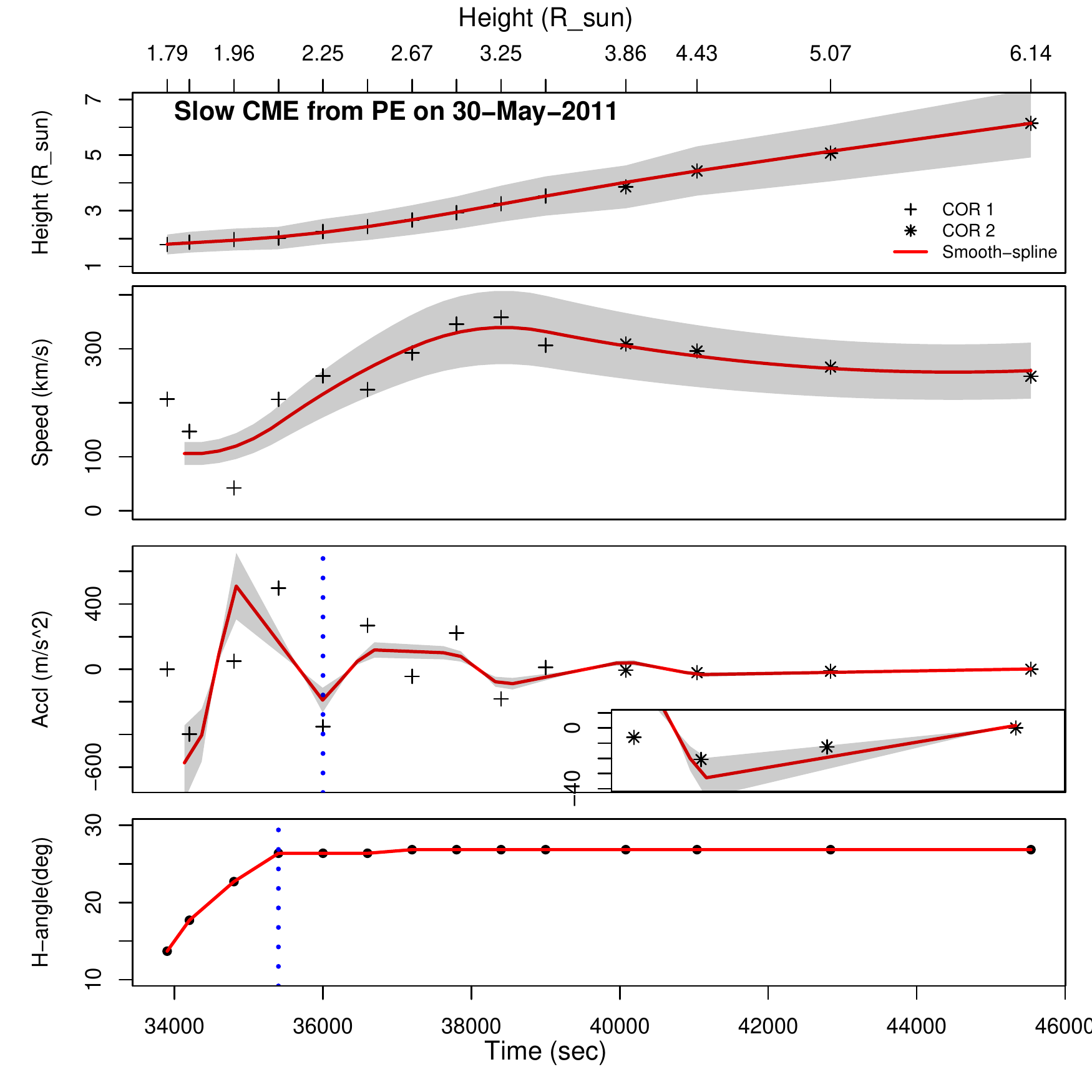}{0.52\textwidth}{(b)}
        }
\gridline{\fig{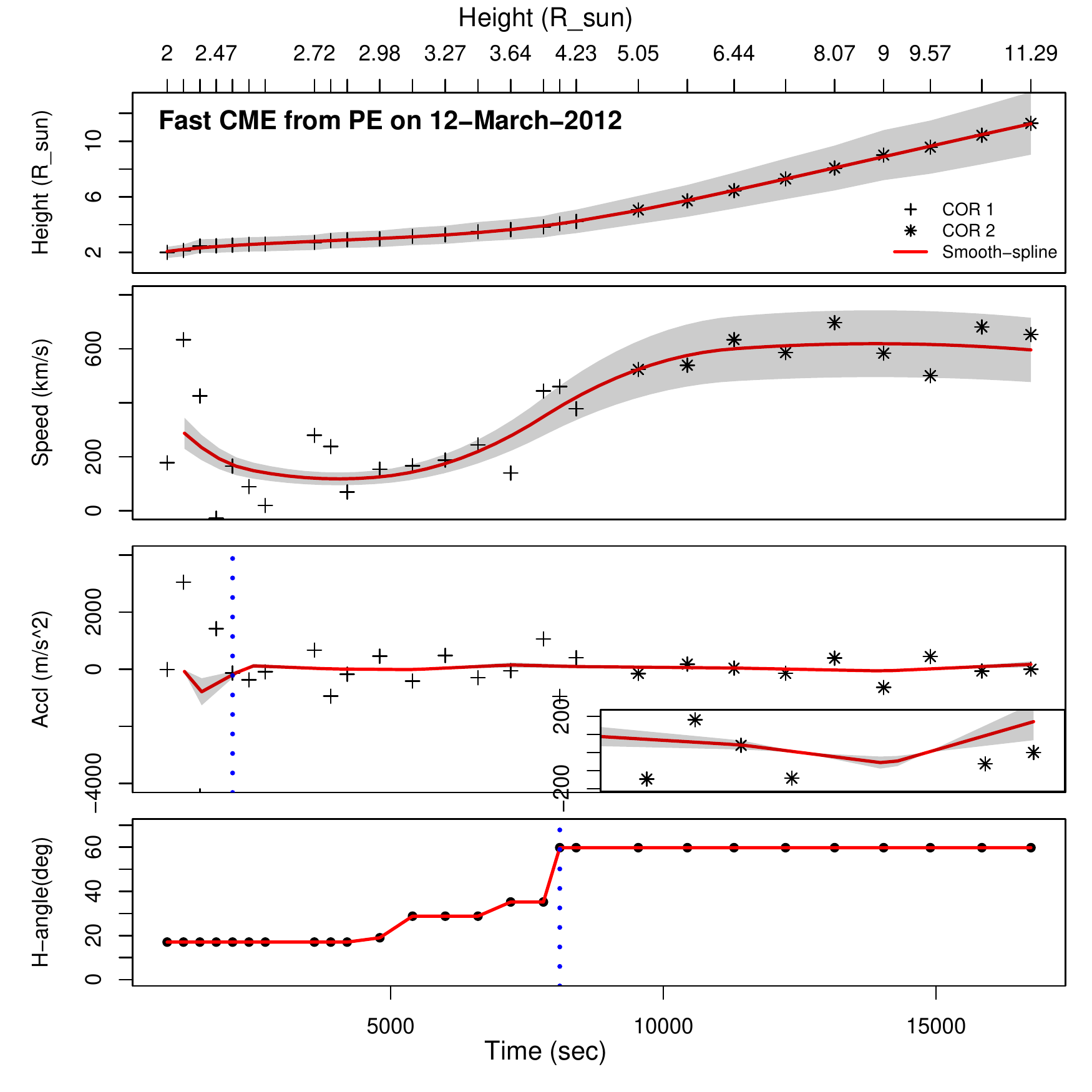}{0.52\textwidth}{(c)}
          \fig{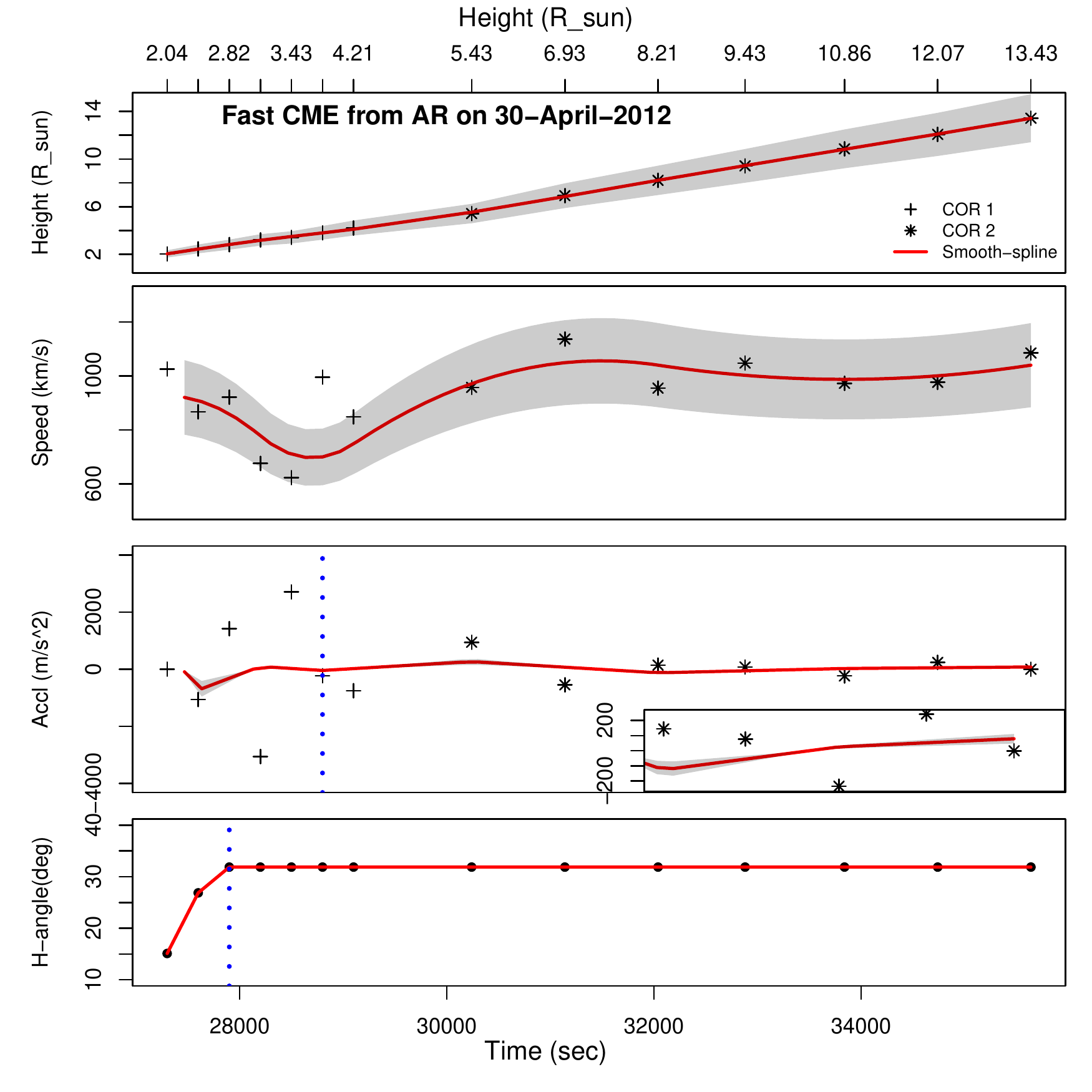}{0.52\textwidth}{(d)}
          }
\caption{3D kinematic profiles of slow CMEs in (a) and (b), and fast CMEs in (c) and (d). In every panel, the top figure shows the height-time plot derived from COR-1 and COR-2 data using the GCS model. The solid red line represents the smooth-spline fit to the height-time data. The second and third plots show the CME speed and acceleration derived by the numerical differentiation of (a) and (b), respectively. The uncertainty in the fitted model is shown in the grey shaded region.An inset with a zoom into the residual acceleration phase is provided for each event in the bottom corner of the acceleration plots. The axis scale in x coincides with the x axis at the bottom.}
\label{kinem}
\end{figure*}

Since we aim to understand the 3D evolution of CMEs in the inner and outer corona, in this section we explore the 3D kinematic profiles of them. Of the three forces (Lorentz force, viscous drag force and gravitational force), that govern the dynamics of CMEs, since Lorentz force is the main driving force in the initial eruption process \citep{Kliem_2014,Isenberg_2007} that leads to the initial translation and expansion of the CME \citep{Subramanian_2014}, we connect here the true width evolution of CMEs to their 3D kinematic profiles.  
Figure \ref{kinem}, shows the kinematic profiles of four different CMEs out of which (a) and (b) are two slow CMEs and the other two are fast ones based on their speeds recorded in CDAW catalogue. Every panel further contain four subplots. From the top, the first panel shows the height-time curve of the CMEs derived using GCS model fitting on COR-1 and COR-2. The height-time data is fitted with a cubic smooth spline which has an additional advantage of choosing the number and position of nodes. This provides the advantage of capturing the different stages of evolution in the kinematic profiles such as the initial impulsive phase, the gradual acceleration phase because the acceleration of a CME does not remain constant but changes with height \citep{2007SoPh..241...85V}. For the fitting, we choose on average from six to ten knots to characterise the spline. A similar fitting procedure was adopted by \citet{2013SoPh..286..241M} and \citet{2011ApJ...738..191B}. The uncertainty in fitting due to the uncertainty in the fitted model parameters (refer Section \ref{sec2}) is shown by the light grey shaded region around the fitted spline in red (Figure \ref{kinem}).  We use only white light data for the study of kinematics so as to ensure that we are tracking the same physical feature in the FOV of different instruments. The second and third panels from the top show the variation of the speed and acceleration with time derived by numerical differentiation of the height and velocity, respectively. To understand the nature of variation of the width of CMEs, in the bottom panel, we plot the evolution of the half-angle of the CME, which signifies the width of the CME. We also show the height axis on top for a better understanding. Figure \ref{kinem}(a) shows a slow CME coming from AP on April 2, 2011. The second and third panels show that the CME experienced an initial impulsive acceleration which accelerated the CME to a high speed and then it got decelerated, with the speed eventually slowing down and becoming constant at a smaller speed. Such increase and then decrease in speed of a CME hints towards the drag experienced by the CME due to the solar wind, which reduces the speed to a lesser value and the CME propagates with that constant speed as reported by \citet{nat2000}. The acceleration also shows a similar nature of variation, with a sudden initial impulsive phase and then almost vanishes for the later part of the propagation phase. The expansion of the CME shown in the fourth panel shows that the CME expand during its initial evolution and then propagated with a constant width beyond 2.57 R$_{\odot}$.  Figure \ref{kinem}(b) shows a similar behaviour, but the speed after attaining a higher value, became constant at that value, without showing much signs of further deceleration. A similar behaviour is seen for panels (c) and (d) of Figure~\ref{kinem}. Similar profiles were also reported by \citet{2011ApJ...738..191B}, in this work we report on true speed and acceleration profiles.\\

 It is understood that the viscous drag force dominates over other forces at the higher heights \citep{Sachdeva_2015}, and the CME is expected to have a constant speed profile with little or no acceleration \citep[also refer][]{article}. Thus, we plot a blue dotted vertical line on the acceleration plot corresponding to the height where the initial impulsive acceleration vanishes. We also draw a similar vertical line on the width evolution plot, where the width is becoming constant there upon. A comparison of these two heights (the height at which impulsive acceleration vanishes and the height at which the width becomes constant) will tell us the height till which the Lorentz force leaves its imprint on the kinematic profiles of the evolution of CME from the inner corona, to the heliosphere. For event (a), the acceleration drops down at 3.28 R$_{\odot}$, while the width becomes constant at 2.57R$_{\odot}$. Similarly, the respective heights for events (b), (c) and (d) are (2.25 R$_{\odot}$, 2.02 R$_{\odot}$), (2.52 R$_{\odot}$, 4.06 R$_{\odot}$) and (3.81 R$_{\odot}$, 2.82 R$_{\odot}$) respectively. We note that the heights of vanishing impulsive acceleration and expansion are similar for events (a), (b) and (d), while there is a reasonable difference in the two heights for event (c), where we find that despite the initial acceleration ceasing at 2.52R$_{\odot}$, the CMEs show expansion till 4.06R$_{\odot}$. We do not understand the reason yet for the difference in the heights in event (c) and a detailed analysis is needed for a better understanding.

 Thus, in the 3D kinematic profiles, we find that CMEs first experience impulsive acceleration, followed by a gradual acceleration and finally a little or no acceleration, confirming early works from a single vantage point \citep{Zhang2001, Zhang_2004,2011ApJ...738..191B}. It is also evident that the average values of different kinematic parameters like speed, width, acceleration fail to provide insights about the complete nature of variation and hence can be misleading. Motivated by the varied width evolution, and the connection between width expansion and vanishing impulsive acceleration, we study the evolution of true width of CMEs and the impact of Lorentz force on the kinematics.

\begin{figure*}[h]
\gridline{\fig{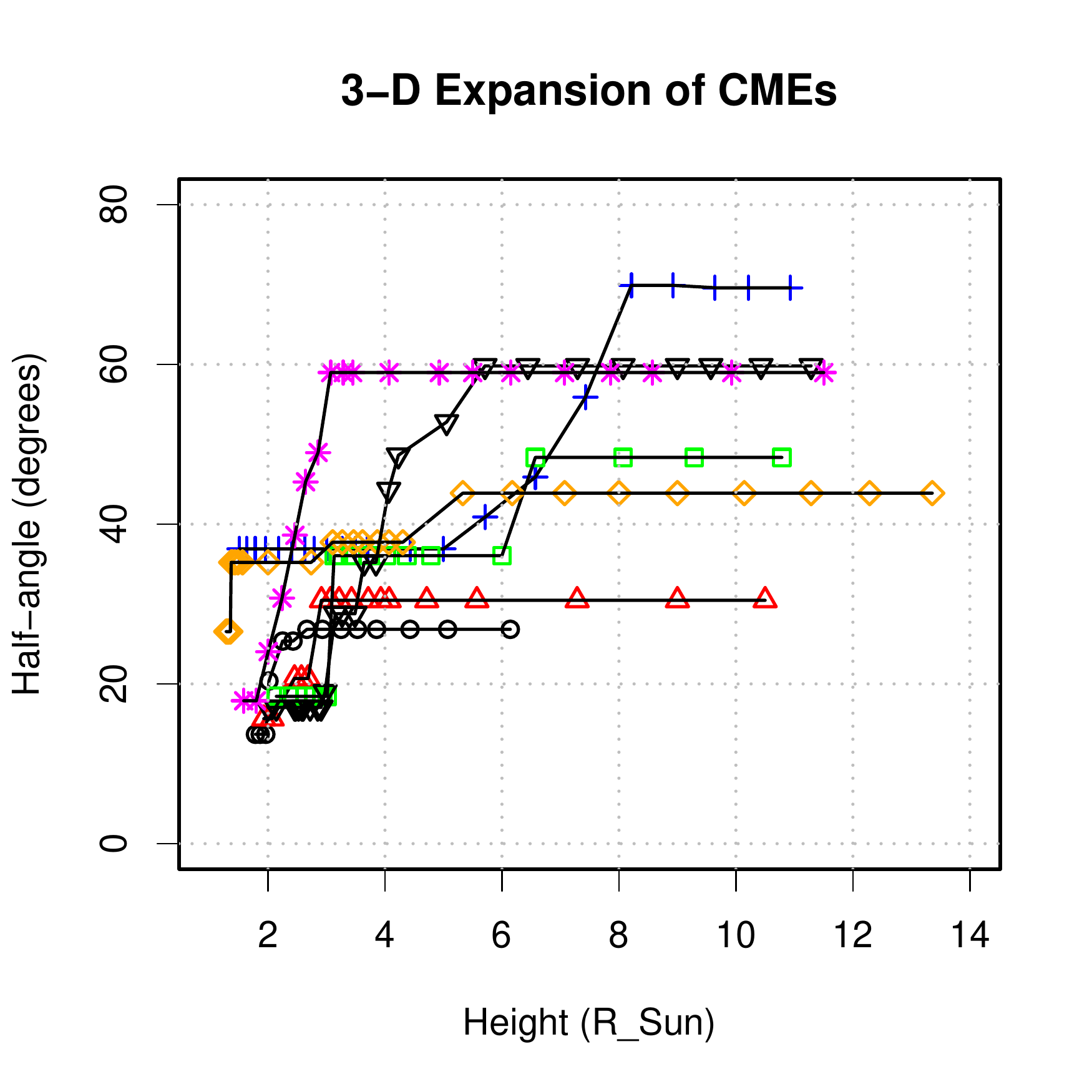}{0.45\textwidth}{(a)}
          \fig{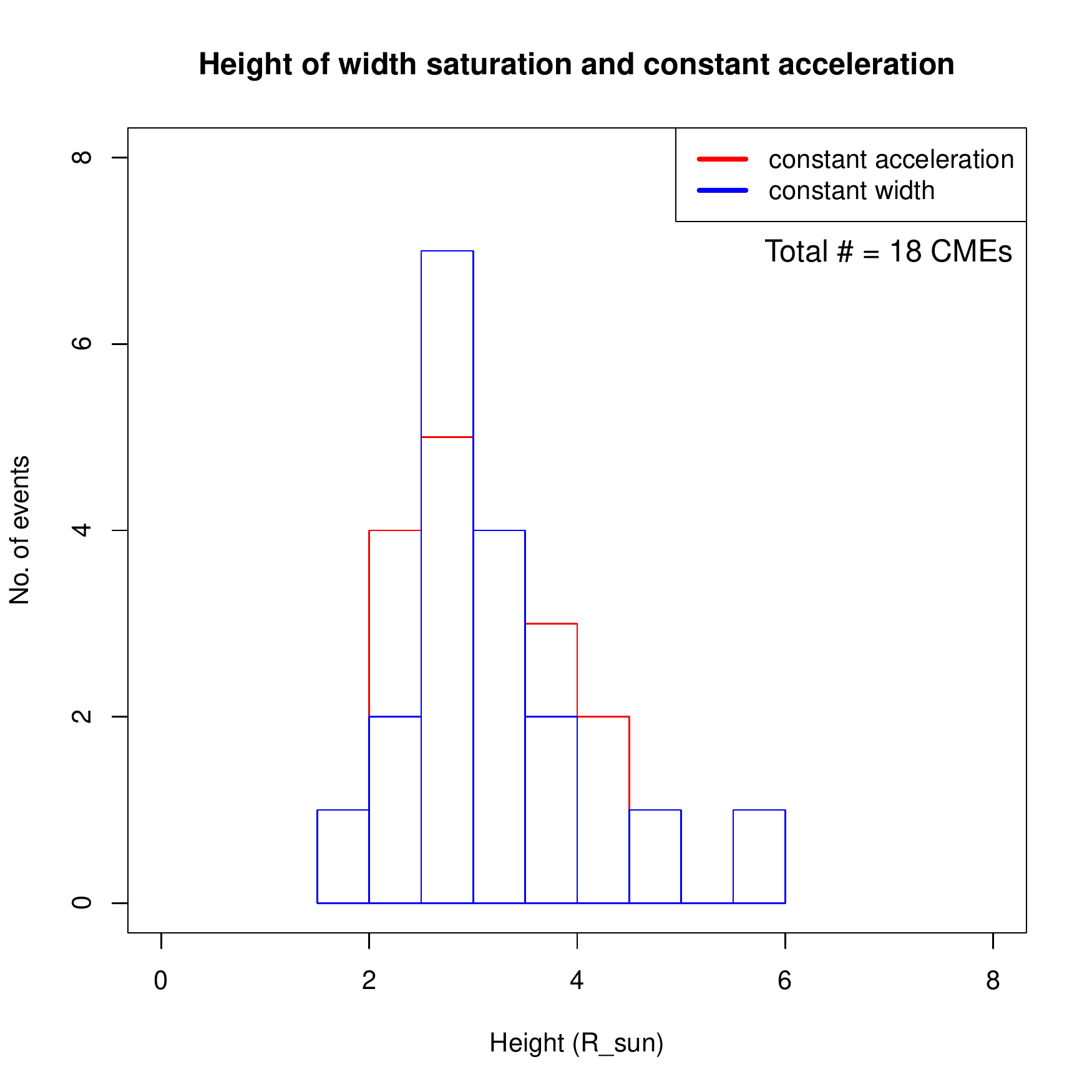}{0.45\textwidth}{(b)}
         }
\caption{(a): Variation of the half-angle of CMEs with height. Only those events are showed which showed expansion by greater than 10 degrees. Different colors correspond to different events. (b): Distribution of the height of the saturation of the width and the height where impulsive acceleration vanishes.}
\label{exp}
\end{figure*}
We study the evolution of true widths of CMEs in inner and outer corona and then further connect the true widths of CMEs to their true accelerations to have a better understanding of the impact of Lorentz force on their 3D evolution close to their initiation heights. Using the fitted GCS model parameters, we select the CMEs which showed a change in their half-angle as they propagated outwards. In Figure \ref{exp}(a), we plot the change in half-angle of the CMEs which show a change of more than 10 degrees of their initial half-angle ($\alpha$). A change in $\alpha$ exhibits expansion of CMEs \citep{2009SoPh..256..111T} and we found that in $~27$ $\%$ (16 out of 59)  cases, CMEs showed true expansion ($~$12 $\%$ (7 out of 59) of those showed more than 10 degrees of expansion, refer Figure \ref{exp}(a)). We see that some CMEs show initial rapid increase in width and then a subsequent saturation, whereas, some cases show a very gradual increase in the width and then it becoming constant. The expansion of the CMEs in most of the cases is found to saturate before 4 R$_{\odot}$, whereas, in a few cases the CMEs keep on expanding till 8 R$_{\odot}$. Thus we find a wide range of heights where the width of the CME attains a constant value. Since, the Lorentz force leads to the impulsive acceleration and their expansion in the inner corona, we selected the events which showed impulsive acceleration and 3D expansion in their kinematic profiles (refer Figure \ref{kinem}). In Figure \ref{exp}(b), we plot the histogram distribution of the heights where the widths become constant (in blue) and the distribution of heights where the impulsive acceleration of the CME vanishes (in red). We find that both the distributions have a well overlapping height ranges, while the distribution of heights of constant width has a wider spread than that of the heights of vanishing impulsive acceleration. We find that for most of the CMEs, the width became constant in the height range of $2.5-3$ R$_{\odot}$. This is in agreement with the results of \citet{2020A&A...635A.100C} who  reported that the width of the CMEs changed considerably below 3 R$_{\odot}$. 
 We also note from panel (b) that the mode of both the distributions fall in the overlapping height range of $2.5-3$ R$_{\odot}$. This tells us that for most events, the vanishing of initial impulsive acceleration and the constancy of width happens in this height range. Thus, this is the height range till which the impact of Lorentz force remains evident in the kinematic profiles on the initial evolution of CMEs. This is also in agreement with the results of \citet{2017SoPh..292..118S} who reported that Lorentz forces peaked in the height range of $1.65-2.45$ R$_{\odot}$ in the evolution of CMEs.

\subsection{Affinity for the equator - 3D Deflection of CMEs}\label{sec5}

\begin{figure*}[!ht]
\gridline{\fig{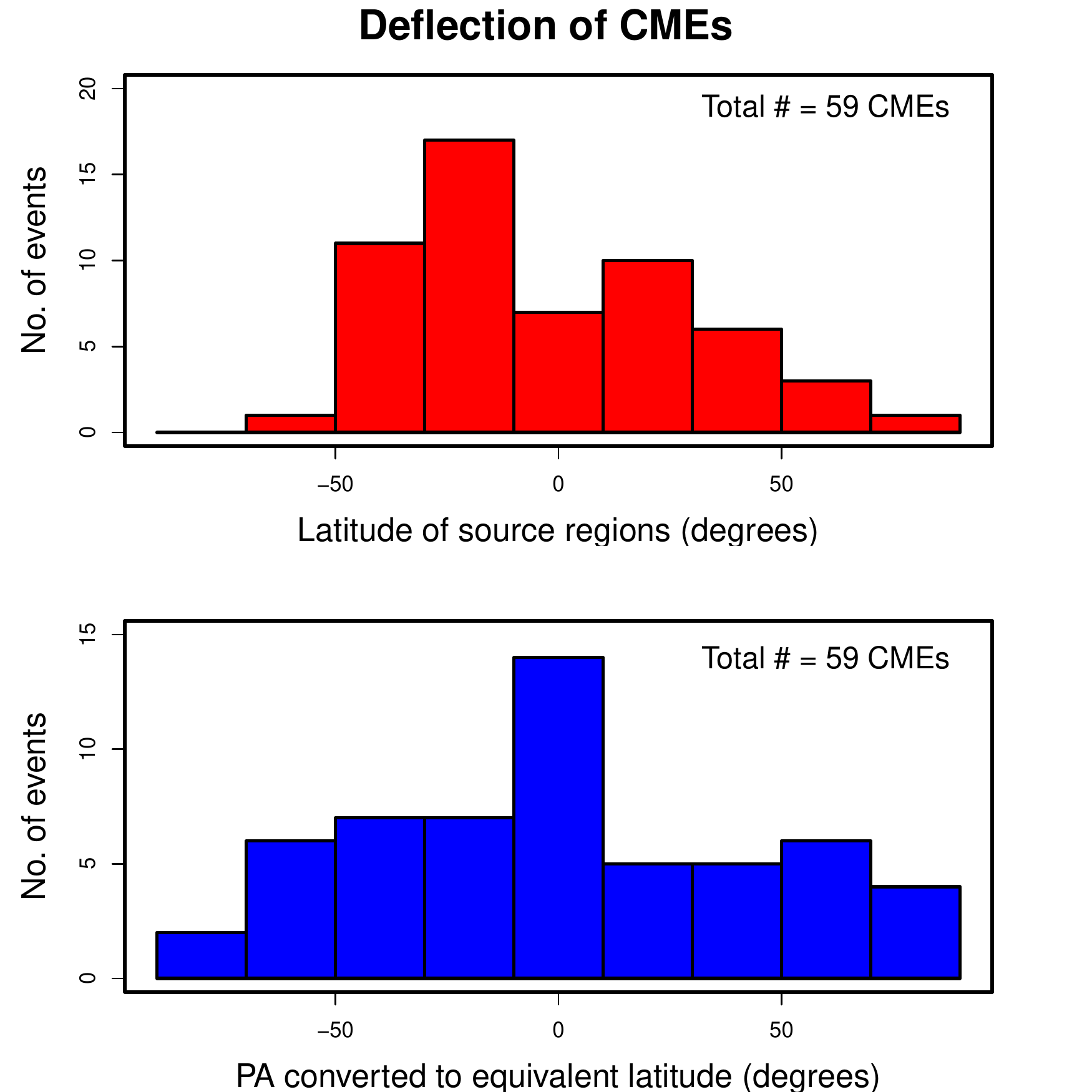}{0.45\textwidth}{(a)}
          \fig{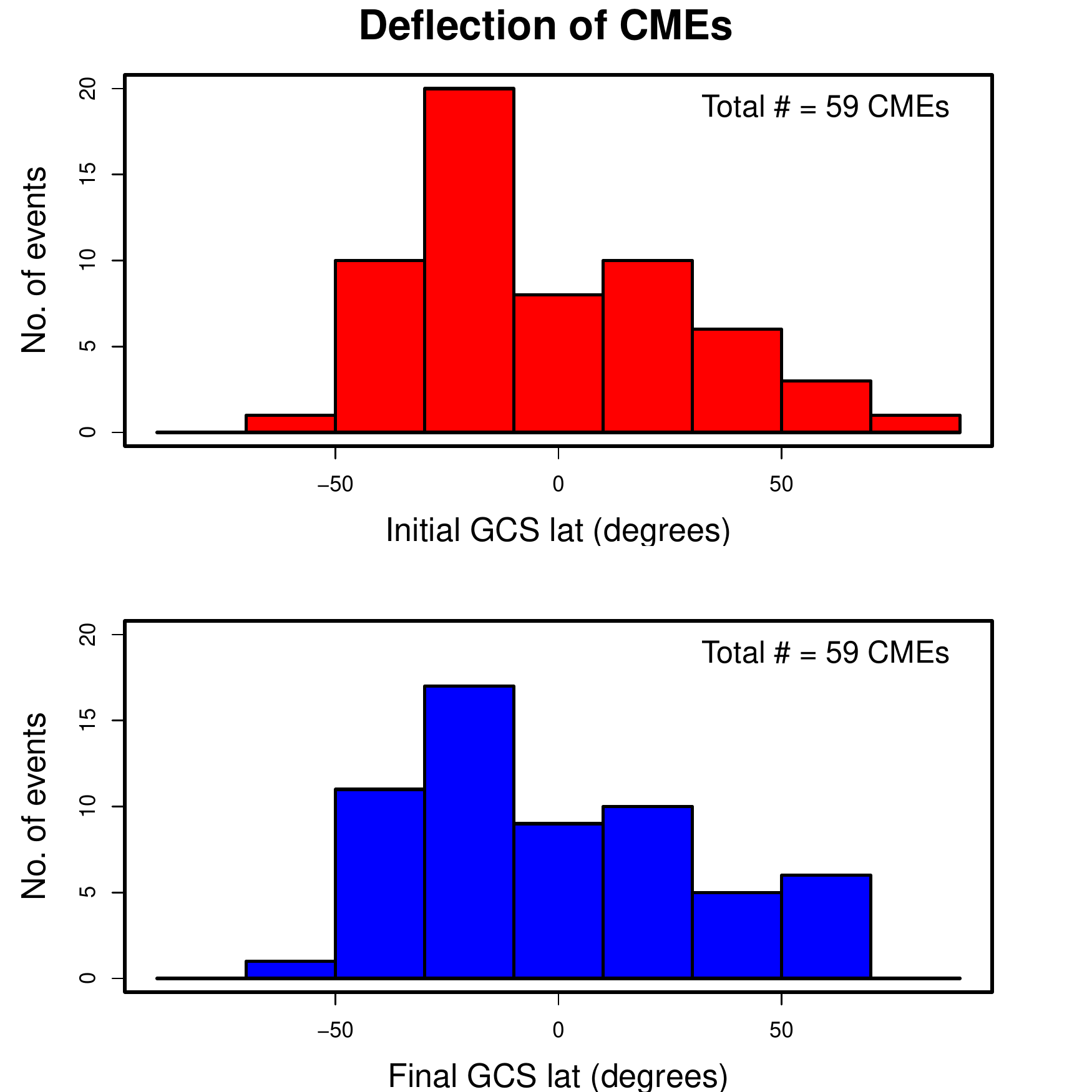}{0.45\textwidth}{(b)}
          }
\gridline{\fig{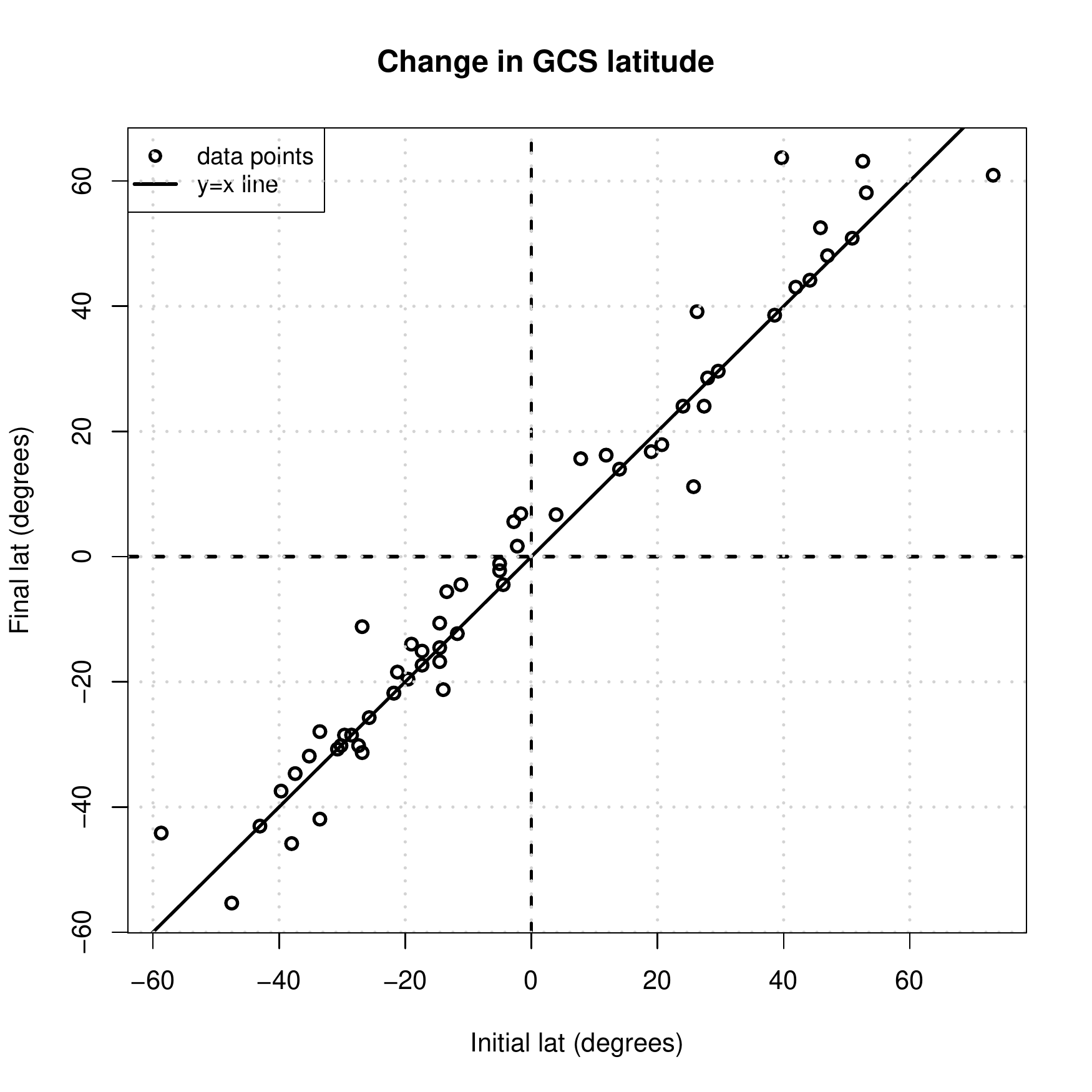}{0.45\textwidth}{(c)}
          \fig{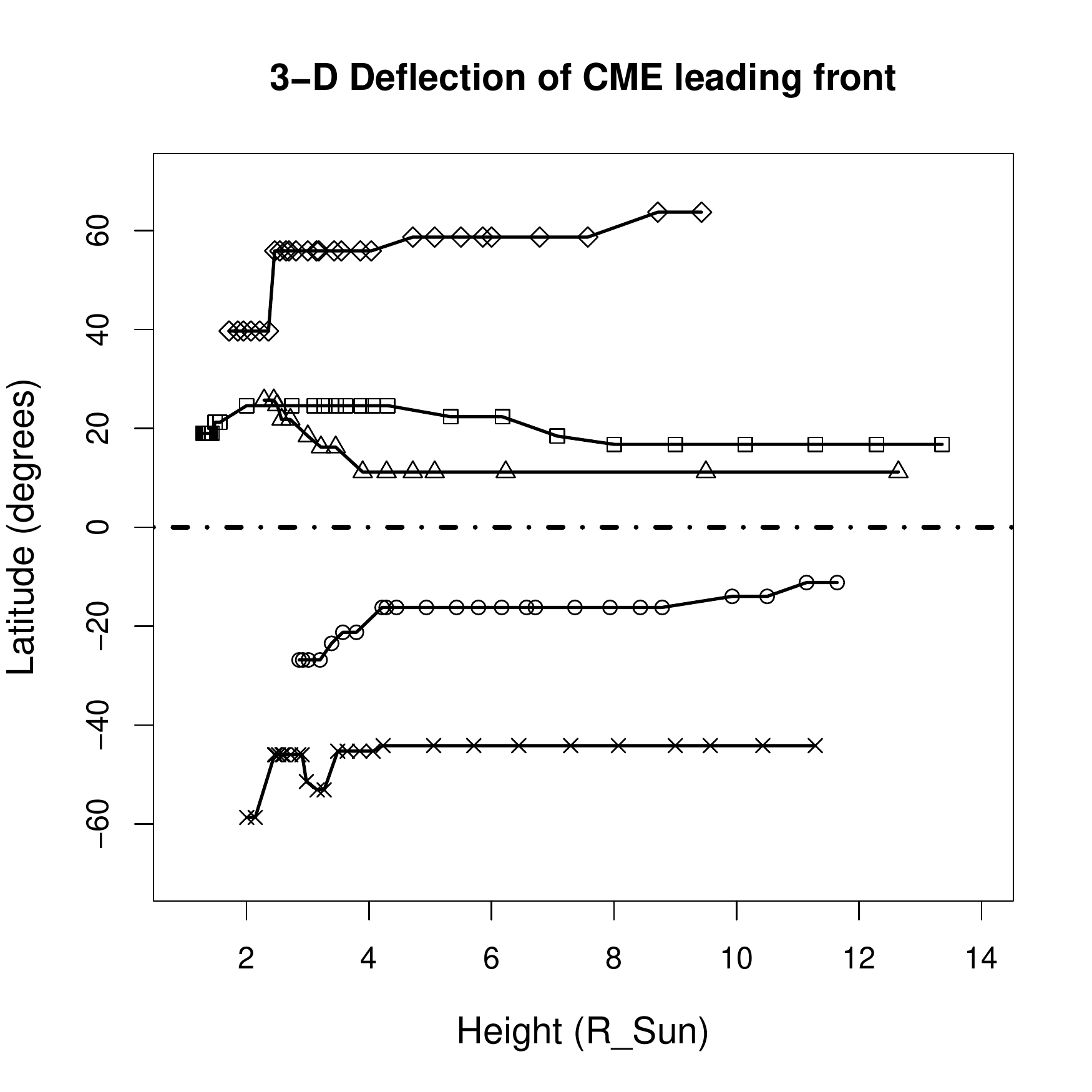}{0.45\textwidth}{(d)}
          }
\caption{(a) Top panel: The latitudinal distribution of the locations of source regions of CMEs. Bottom panel: Distribution of the PA equivalent latitude of the CME leading front.  (b) Top panel: Distribution of the initial GCS latitude. Bottom panel: Distribution of the final GCS latitude. (c) Plot of final versus initial GCS latitude, with the solid line being the boundary where both the values are equal. The dashed line represents the zero latitude (i.e. equator) (d) Change of latitude of the CME with height estimated from the GCS model. only those cases where the deflection was at least more than 10 degrees are shown. The different data points denote different events - 2011-01-24 (circle), 2011-06-20 (triangle), 2012-02-24 (square), 2012-03-12 (cross) and 2012-10-12 (rhombus). }
\label{defl}
\end{figure*}

We investigate whether or not CMEs show any signatures of deflection from a radial path of propagation.   We first look into the latitude distribution (refer top panel of Figure \ref{defl}(a)) of the source regions of the CMEs. We can see that it shows a bi-modal distribution with the peaks of the distribution lying in the latitude range of $\pm$ $10^{\circ}-30^{\circ}$, which led us to believe that most of the source regions were associated with ARs \citep{2018SoPh..293...60M}, which in our case are either ARs or APs. In the bottom panel of the same figure, we convert the central Position Angle (PA) of the leading front of the CME as given in the CDAW catalogue, to its equivalent latitude and again plot the distribution of the PA equivalent latitude. Unlike the case for the bi-modal distribution of source location latitudes, for the CMEs we find a broad distribution with a distinct peak around the equator. This tells us that a majority of the CMEs while ejecting, got deflected towards the equator. Similar signature was also reported by \citet{Gopalswamy_2003} as they found CMEs originating from prominences got predominantly deflected towards equator from the latitude distribution of prominences and that of the CMEs. However, the latitude distribution in that study suffers from the projection in the plane of the sky. To check whether the deflections (from Figure \ref{defl}(a)) were actual deflections suffered by the CMEs, we look into the GCS latitude of all the events. We plot in (b) the distribution of initial GCS latitude (top), and final GCS latitude (bottom). In this case we do not find such change in the latitude distribution as we find in (a). We find that the distribution remains bi-modal. To further look into this, we plot the final versus initial GCS latitude. The solid line is the boundary where both the initial and final latitudes are equal, and hence the points lying on this line denotes CMEs which did not get deflected. The dotted line marks the equator. The data points lying to the left of the vertical dotted line denote CMEs with initial latitude in Southern hemisphere. In this region, the points lying above the solid line shows events for which the final latitude is greater than the initial latitude, and thus denote equator-ward deflection, and vice versa in the region to the right of the vertical dotted line. As we see from the figure, we do find that most of the CMEs which suffer deflections, get deflected towards the equator. But in most of these cases, the angle of deflections are small, and hence a distribution of GCS latitude does not show a single mode around the equator. From this, we find that the apparent deflection as found from the latitude v/s PA equivalent latitude plot is not a conclusive proof of deflection of CMEs, as the actual deflection suffered (as measured from gcs lat) is much lesser. Due to the uncertainty in the model fitting ($\sim 20$\%), in Figure \ref{defl}(d), we plot the variation of GCS latitude with height for only those cases which showed deflection greater than 10 degrees. In this case too, we find that most of the CMEs get deflected towards the equator irrespective of their location being in the northern or southern hemisphere. In this context, \citet{nat}, \citet{Kahler_2012} reported that CMEs can get get deflected from their initial path  when they get ejected near a coronal hole. In our sample of events, using the data from AIA and EUVI, we found that for all the events that show large deflections ($>10$ degrees) coronal hole was present near their location. We used the JHelioviewer software \citep{jhelio2,jhelio} and the data from SDO/AIA 193 \AA\ and STEREO/EUVI 195 \AA\ and identified the location of the coronal holes near the source regions of the deflected CMEs. With the coordinates of the coronal holes provided by the Jhelioviewer, we found that on an average the coronal holes are located within $17^{\circ}$ of latitude of the eruption. \citet{Kay_2015} reported that such interaction of CMEs with the open field lines of coronal holes guides the CMEs towards the heliospheric current sheet. Further, we also found that five events plotted in Figure \ref{defl}(a) have either active or quiet prominences (or filaments) as their source regions. This is also supported by the reports on non-radial ejections of CMEs from prominences with majority of them getting deflected towards the equator by \citet{Gopalswamy_2003}. 

Thus we find that, first, the CMEs propagate in a non-radial path from their source region, and most of them got deflected towards the equator . This initial deflection towards the equator was also sustained in some cases (refer Figure \ref{defl}(a) and \ref{defl}(b)) where the CMEs continued getting deflected further towards the equator as it propagated into the outer corona. We found true deflection in $~31$ $\%$ (18 out of 59) cases ($~$9 $\%$ (5 out of 59) of those showed more than 10 degrees of deflection). We found the average height till which CMEs got deflected around 3.35 R$_{\odot}$. 
These deflections are also very crucial from the perspective that they provide an indirect evidence of the CMEs exhibiting interactions with ambient coronal structures like coronal holes and can largely affect height-time measurement in 2-D images, giving spurious results. Thus, although from the latitude v/s PA equivalent latitude plot (Figure \ref{defl}(a)), we find the CMEs to be apparently getting deflected towards the equator, from the distribution of GCS latitudes (Figure \ref{defl}(b)), we do not find such strong signatures of equator-ward deflection. Although, we still find from Figure \ref{defl}(c) that majority of the deflected CMEs get deflected towards the equator, yet for only for 5  cases we find the angle of deflection is greater than 10 degrees. Thus we conclude that from our analysis, we do not find a significant statistical deflection, as expected from earlier works \citep{Gopalswamy_2003}, and thus Figure~\ref{defl} (a) does not provide a conclusive evidence of deflection of CMEs, as the numbers largely suffer from projection effects. Further extending this study on a larger data set with actual 3-D values will help establish our conclusions better. \\

\subsection{Comparison of average projected and true Speeds} \label{compar}

\begin{figure*}[h]
\gridline{\fig{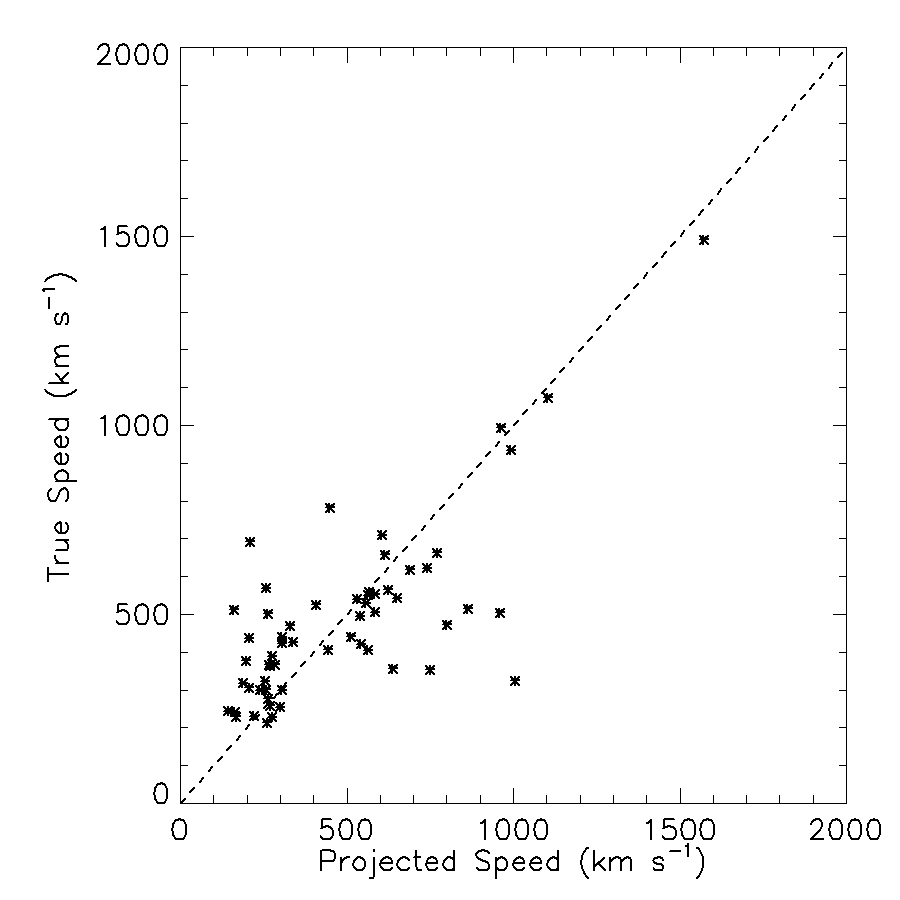}{0.5\textwidth}{}
         }
\caption{A comparison between the average projected 2-D speeds as recorded in the CDAW catalogue and the average 3D speeds as calculated using the GCS model. The dashed line is the boundary where both the 2-D and 3D speeds are equal.} 
\label{vel_comp}
\end{figure*}

 Since our events are selected from the CDAW catalogue, it is worthwhile to compare results that we have obtained applying GCS model with those already catalogued using 2-D studies.  We first find the average 3D speed of all the CMEs as estimated using the GCS model. We find a wide range in the average true speeds from 213 km~s$^{-1}$ to 1492 km~s$^{-1}$. The CMEs were segregated into slow and fast based on their average 2-D speed as quoted in the CDAW catalogue. In Figure \ref{vel_comp} we plot the average true speed versus the average projected speed for slow and fast CMEs. Each data point represent the average speed of one single event. The dashed line is the boundary where both the projected and true speeds are equal. In case of the slow CMEs, we find that, almost all the true speeds are higher than the projected speeds, which is expected. We thus find that a good fraction ($\sim52\%$) (15 out of 29) of the slow CMEs ($< 400$ km~s$^{-1}$ in CDAW catalogue) have higher true speeds. In the case of fast CMEs, we find a good fraction ($\sim 63\%$) (19 out of 30) of the events showing 2-D speeds higher than 3D speeds. This contradicts our intuition. \citet{2013AGUFMSH31B2034S} reported several reasons for this discrepancy. The GCS model does a 3D fitting of the leading front of the CME, whereas speeds in the CDAW catalogue is calculated by tracking the part of the leading edge which move with the highest speed (which may be a part of the shock front of the CME). We also note that the projected speeds are higher than the true speeds mostly for the fast CMEs which are capable of driving shocks. We also found that all these fast CMEs with projected speeds higher than true speeds were marked as partial-halo events in the CDAW catalogue. In this context, \citet{2009CEAB...33..115G} and \citet{2013JGRA..118.6858S} reported that the lateral expansion speed might be greater than the radial propagation speed, and in the projected image, especially for halo or partial halo CMEs, it is difficult to distinguish between the two. Thus, for such events which are halo or partial-halo with respect to one viewing point, it is important to study their kinematics by performing a 3D reconstruction with images taken from more than one vantage point and then find their true propagation speed and not their lateral expansion speed. Further, different values of solar wind speed at different latitudes can have different imprints on different parts of the CME leading edge, which will also influence the results of CDAW catalogue (as they track one single point on the leading edge). But the GCS model fits the leading front completely and hence this effect will be much lesser. We also note from Figure \ref{kinem} that the average values fail to provide the complete picture of the kinematics and hence can be misleading.

\subsection{Distribution of Peak Speeds and Accelerations}

\begin{figure*}[h]
\gridline{\fig{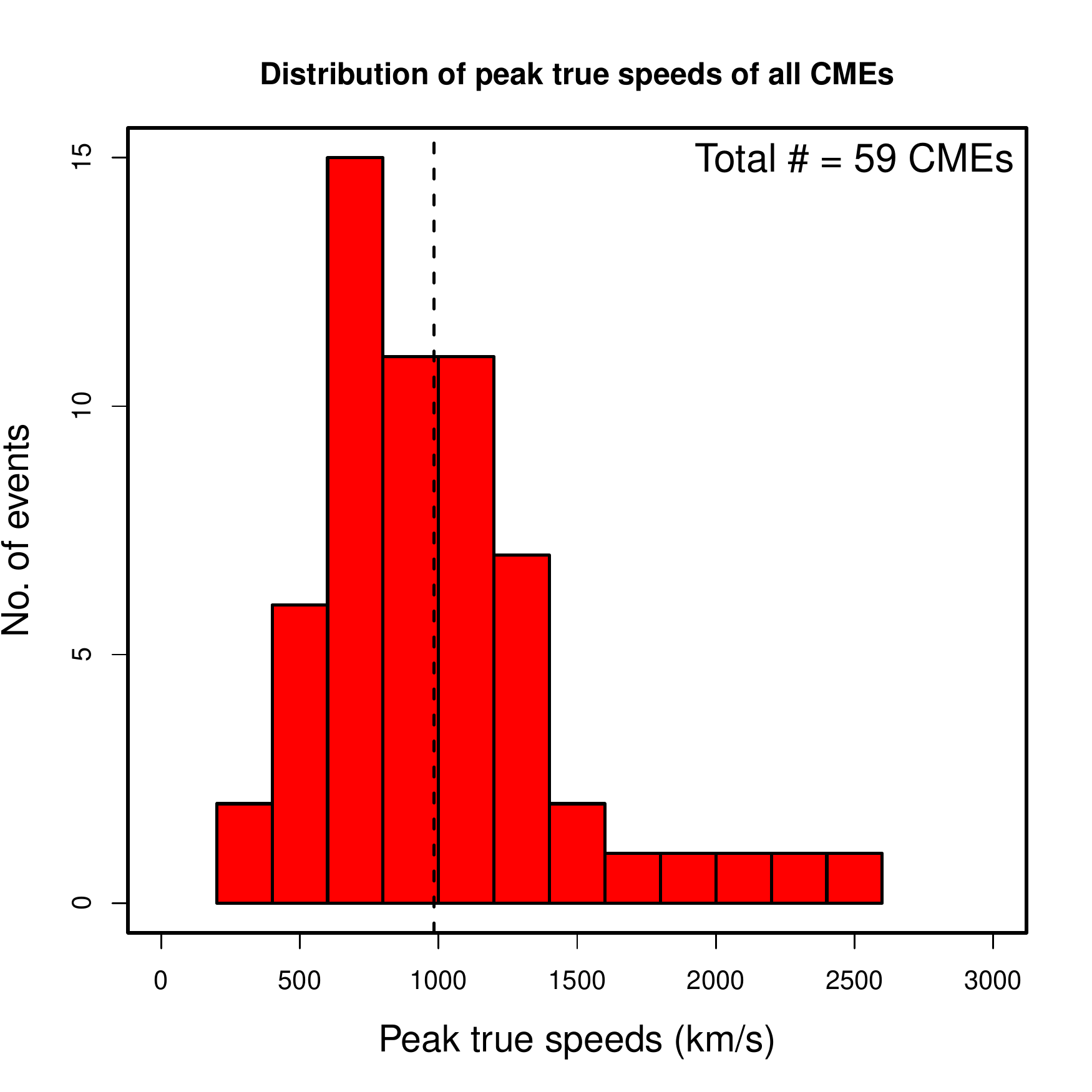}{0.50\textwidth}{(a)}
          \fig{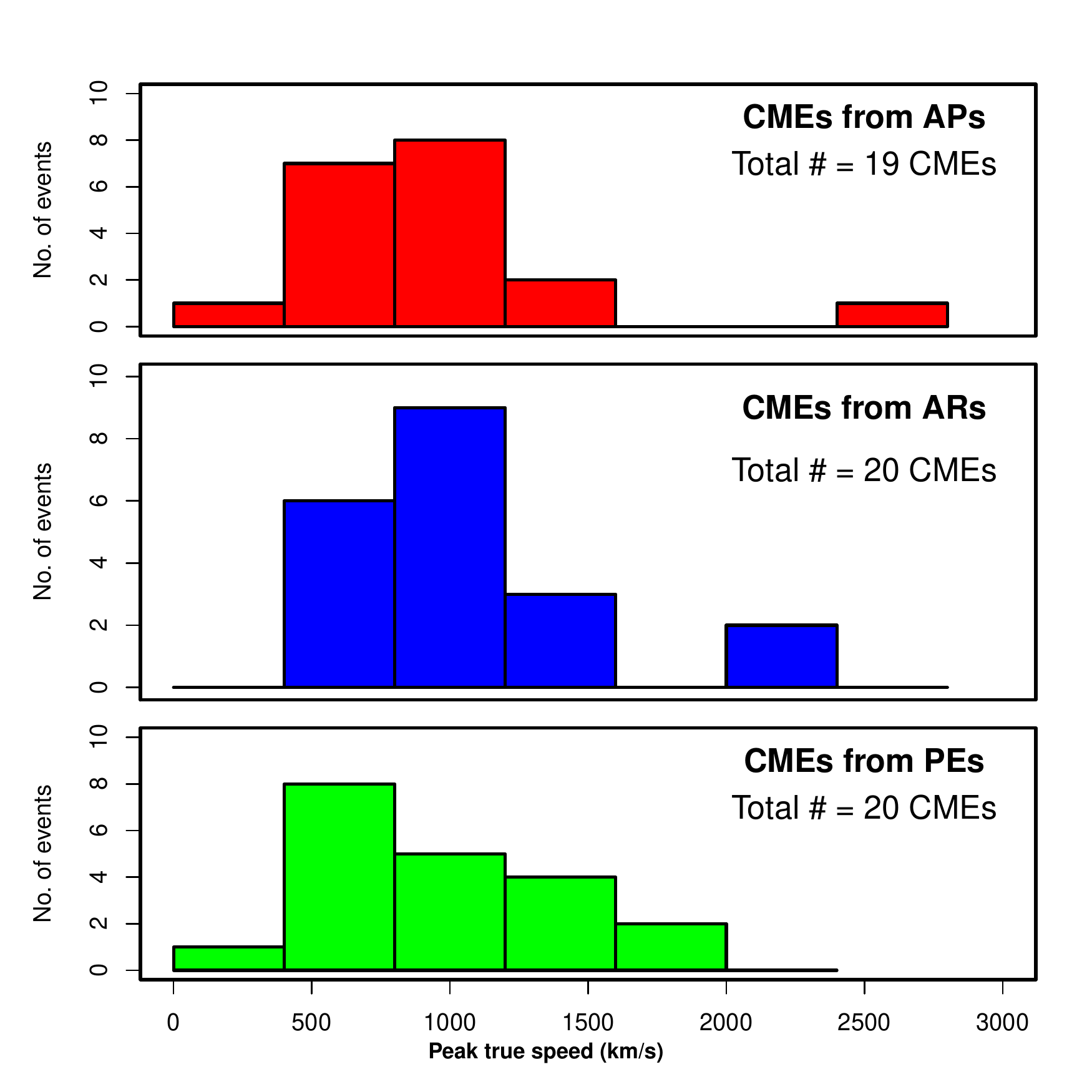}{0.50\textwidth}{(b)}
         }
\caption{(a) Distribution of peak speeds of all the CMEs. The dashed line denotes the mean value}. (b) shows the contribution of peak speeds of CMEs originating from APs, ARs and PEs to (a). 
\label{peak_vel}
\end{figure*}

\begin{figure*}[h]
\gridline{\fig{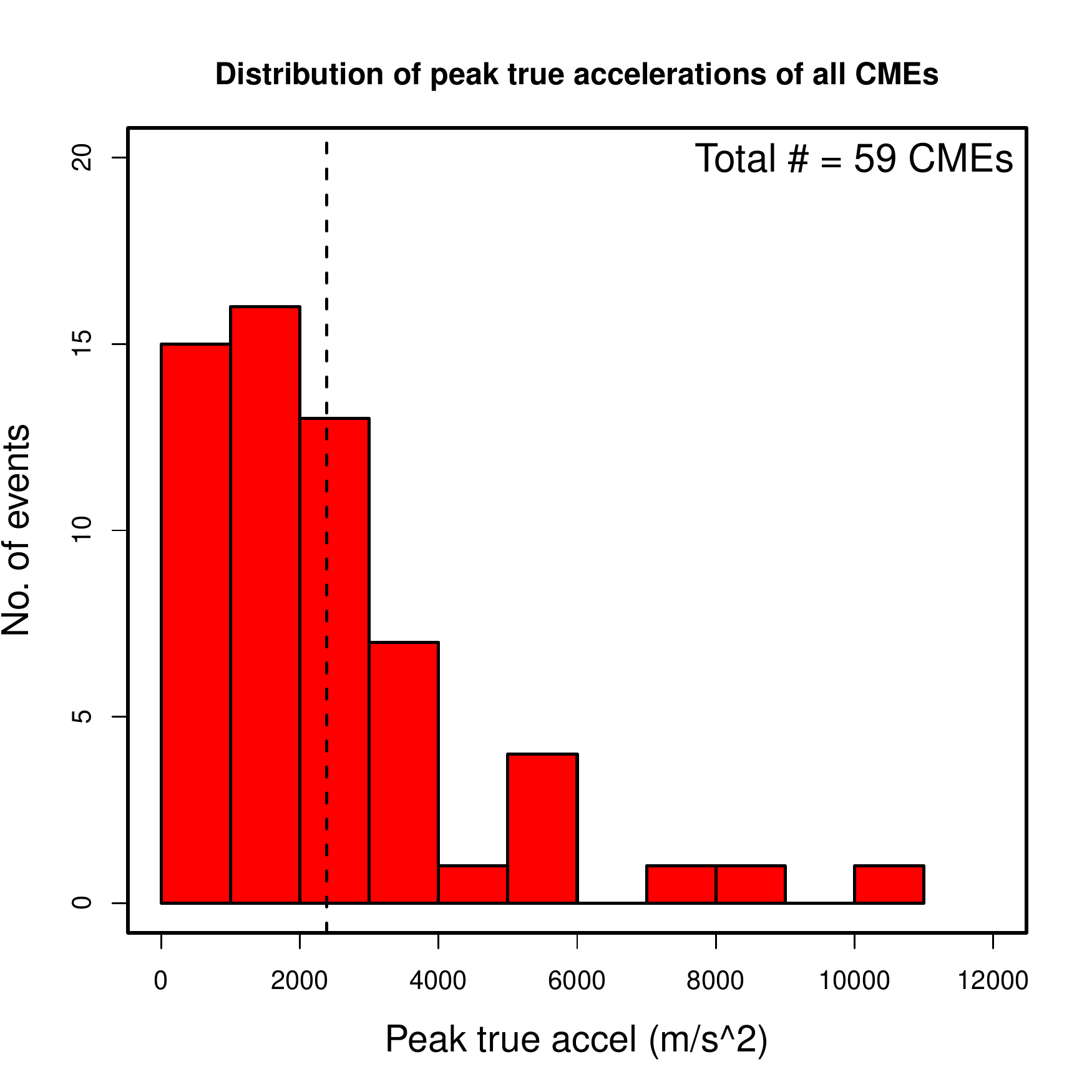}{0.50\textwidth}{(a)}
          \fig{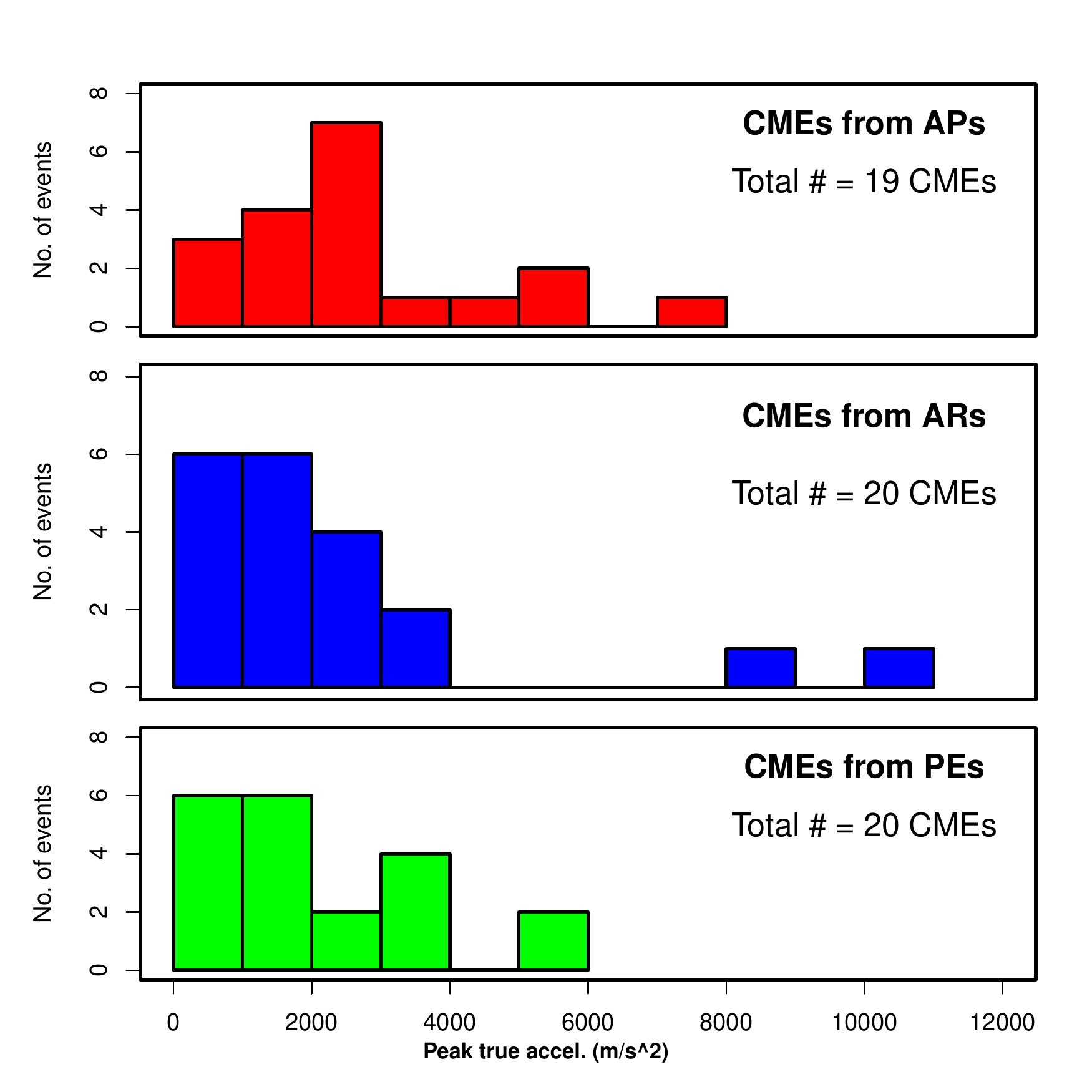}{0.50\textwidth}{(b)}
          }
\caption{Panel (a) shows the distribution of peak accelerations of all the CMEs. The dashed line denotes the mean value}. Panel (b) shows the contribution of peak accelerations of CMEs originating from APs, ARs and PEs to panel (a).
\label{peak_accl}
\end{figure*}

In Figure \ref{peak_vel}(a) we plot the distribution of peak 3D speeds ($V_{peak}$) for all the 59 CMEs studied in this work. We find a wide range for $V_{peak}$ from 396 to 2465 km~s$^{-1}$ with a mean value of 984 km~s$^{-1}$, and the peak of the distribution being in the range 600 to 800 km~s$^{-1}$. \citet{2011ApJ...738..191B} reported a range for $V_{peak}$ from 56 to 1279 km~s$^{-1}$ with a mean value of 526 km~s$^{-1}$ and mode lying in the range 300 to 400 km~s$^{-1}$ for 95 events during 2007 to 2010 which was during the solar minimum of cycle 23. \citet{2007SoPh..241...85V} studied 22 events during 2002 to 2005 which was the maximum and decay phase of cycle 23 and found $V_{peak}$ to range between 365 to 2775 km~s$^{-1}$ with a mean value of 940 km~s$^{-1}$. We find our results are similar to the results of \citet{2007SoPh..241...85V} and our events were selected during 2007 to 2014 which includes cycle 23 minimum, cycle 24 rising phase and maximum with a majority of the events ($\sim69\%$) coming from cycle 24 maximum, and further, the speeds in our work are the true peak speeds. \citet{2011ApJ...738..191B} reported on a long tail of the distribution of $V_{peak}$ towards high velocities. We also find a similar trend in our result (Figure \ref{peak_vel}(a)) and since we have the information of the source region, we wanted to look at the contribution of the CMEs coming from the different source region to this distribution. In panel (b), we plot the peak speed distribution but for the CMEs segregated on the basis of their source regions. We find that the tail of the distribution towards high velocities is contributed by the CMEs coming from ARs and APs. We also note that the mode of the distribution is the same for CMEs coming from ARs and APs (800 to 1200 km~s$^{-1}$), while for CMEs from PEs it is between  400 to 800 km~s$^{-1}$, which further implies that the CMEs from ARs and APs tend to have higher peak speeds than the ones from PEs. In this regard, we also found that most of the CMEs (42 $\%$) (25 out of 59) reached their true peak speeds below 5 R$_{\odot}$.\\

We also plot the distribution of peak true acceleration ($a_{peak}$) in Figure \ref{peak_accl}(a). We note that while the distribution of $V_{peak}$ increases gradually to a peak and then decreases with a long tail, in case of $a_{peak}$, the distribution falls off with increasing acceleration, but with a similar tail towards higher values. We further find an even wider range for $a_{peak}$ from 176 to 10922 m~s$^{-2}$ with a mean value of 2387 m~s$^{-2}$ and the mode of the distribution lying in the range 1000 to 2000 m~s$^{-2}$. This is again higher than the range obtained by \citet{2011ApJ...738..191B} from 19 to 6781 m~s$^{-2}$ with a mean of 756 m~s$^{-2}$. whereas \citet{2007SoPh..241...85V} reported a higher range from 40 to 7300 m~s$^{-2}$. Again to find the contribution of different source regions to the peak acceleration distribution, we again plot the distribution separately for the three different sources in panel (b). We again find that the high acceleration tail is coming from the CMEs which had either ARs or APs as their source regions. Thus, based on the distribution of peak true speeds and accelerations, we find that the CMEs originating from ARs and APs are more energetic than the ones coming from PEs. This conclusion is based on the speed and acceleration of these CMEs. The injected flux in the flux-rope is responsible for translating the CMEs \citep{Subramanian_2014,article} , and thus, according to our definition of ARs, and APs, CMEs associated with AR and AP might have more flux injection leading to strong accelerations. This could be responsible for large speeds and accelerations. A detailed investigation needs to be done considering the magnetic field of the source region and the injected flux to the CMEs to further quantify this statement.

\subsection{Acceleration duration and acceleration magnitude}
\label{accel_dur_mag}

\begin{figure*}[h]
\gridline{\fig{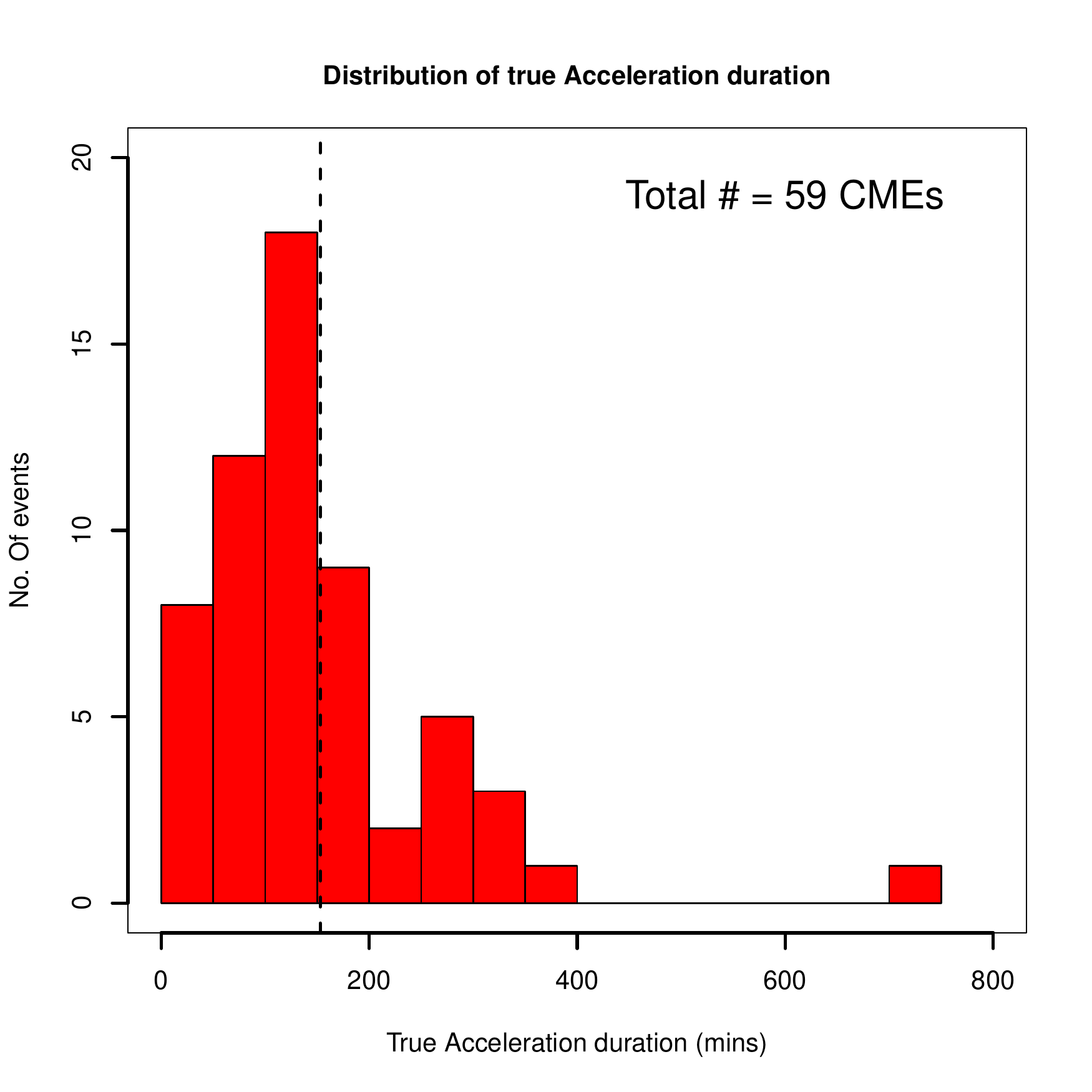}{0.48\textwidth}{(a)}
          \fig{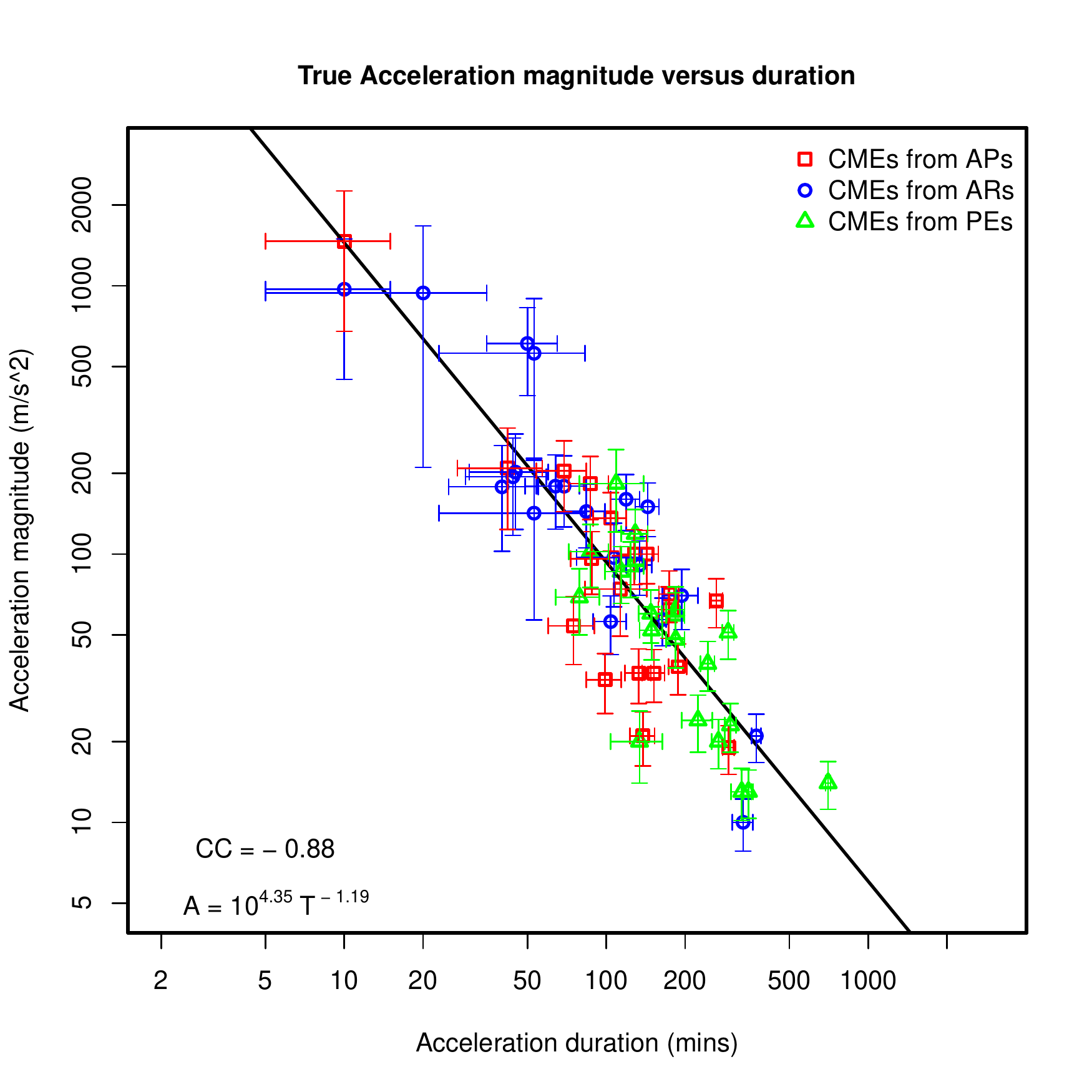}{0.48\textwidth}{(b)}
         }
\caption{(a) Distribution of acceleration duration of all the CMEs. The dashed line denotes the mean value. (b) Acceleration magnitude versus duration of all the CMEs. Different colors indicate CMEs originating from different source regions. }
\label{accl_dur}
\end{figure*}

Since CME events can range from being extremely impulsive to extremely gradual based on their accelerations, a study of the true acceleration duration and acceleration magnitude is important as they greatly affect the speeds of CMEs \citep{Zhang_2006}. Based on the 3D kinematic profiles (refer Figure \ref{kinem}), we were able to track the change in velocity, thereby identifying the onset and ending time, and hence the acceleration duration. In Figure \ref{accl_dur}(a), we plot the distribution of the acceleration duration in minutes. We find a wide range in the acceleration duration from 10 minutes to 703 minutes with the peak of the distribution in the range of 100-150 minutes and the mean duration being 153 minutes. Such wide range in acceleration duration was also reported by \citet{Zhang_2006}, \citet{2007SoPh..241...85V} and  \citet{2011ApJ...738..191B}  with a mean duration of 180, 120 and 44.6 minutes respectively. We find our average acceleration duration to be in agreement with the results of \citet{Zhang_2006} and \citet{2007SoPh..241...85V} while \citet{2011ApJ...738..191B} suggested such a relatively lower value was due to the fact that majority of the events studied were impulsive events. Since the acceleration magnitude is also important for understanding the kinematics, we further found the acceleration magnitudes of all the CMEs. For this we followed a similar procedure as \citet{Zhang_2006}, who reported that the acceleration magnitude is the increase in velocity divided by the acceleration duration. The  uncertainty in the acceleration duration is determined by the observational cadences, and the same for acceleration magnitude is inherited from the uncertainty in the acceleration duration and the velocity change (\citet{Zhang_2006}). In Figure \ref{accl_dur}(b), we plot the acceleration magnitude versus acceleration duration. We find a strong inverse correlation between acceleration magnitude and acceleration duration, that is higher acceleration magnitudes have shorter duration and vice versa. This inverse correlation can be described with the following relation:

\begin{equation}
    log(A) = 4.35 - 1.19 log(T),
\end{equation}

where $A$ is the acceleration magnitude in m~s$^{-2}$ and $T$ is the acceleration duration in minutes. The correlation coefficient is -0.88. Similar inverse correlation was also reported by 
 \citet{Zhang_2006} (power law index -1.09), \citet{2007SoPh..241...85V} (power law index -1.14) and \citet{2011ApJ...738..191B} (power law index -1.09). We note from our results that the scaling relation remains similar in 3D study. This thus shows that impulsive events tend to have stronger acceleration for shorter duration while the gradual events tend to have weaker acceleration for longer duration. We also note from panel (b) that the CMEs coming from PEs (in green data points) are distributed in the lower region of the scatter plot, indicating that they are gradual events with weaker acceleration magnitude with longer acceleration durations.

\section{Summary and Conclusions}
\label{sec:summary}
In this work we study the 3D evolution of Coronal Mass Ejections in the inner and outer corona originating from different source regions, which are Active Regions, Prominence Eruptions and Active Prominences. A total of 59 CMEs were studied, of which 29 CMEs were slow ($<~400$ km~s$^{-1}$) CMEs and 30 CMEs were categorised as fast ($>~400$ km~s$^{-1}$) CMEs based on their recorded 2-D speeds in the CDAW catalogue. The source regions of all these 59 CMEs were identified. This helped us to connect the true kinematics of CMEs in 3D to the source regions they are coming from. Multi-viewpoint coronal observations from the twin spacecraft STEREO A and STEREO B was used for the 3D reconstruction of the CME morphology by using the Graduated Cylindrical Shell (GCS) model. The data from COR-1 and COR-2 was used in tracking the CME for studying its kinematics. Table \ref{table2} shows the GCS fitting parameters of events we studied in this work. The table also lists alongside, the identified source region of these CMEs and their average true and projected speeds. For CMEs which were observed edge-on (that gives an ice cream cone shape appearance) from both the view points, we couldn't determine the half-angle uniquely for them and as a consequence, the tilt-angle for such events are kept zero as it has no meaning when the half-angle cannot be determined (for details refer \citet{2009SoPh..256..111T}). This problem might be fixed, if another vantage point of observation can be used to fit the model to the event. For example, LASCO C2 data can be used to address this issue.  We conclude our main results that we have discussed in different sections of this paper as follows.
\begin{itemize}

\item We studied the true width evolution of CMEs in the inner and outer corona (refer Figure \ref{exp}(a)), and found that in $~28$ $\%$  cases, CMEs showed true expansion ($~$12 $\%$ of those showed more than 10 degrees of expansion). We also found that some of the width profiles showed an initial rapidly increasing phase and then a saturation, while some events showed a gradually rising phase or a late rising phase and then a subsequent saturation. Another important feature is that there was a wide range of heights ($2.5-8$ R$_{\odot}$) till which CMEs showed an increase in its width. Since, Lorentz force drives the initial rapid acceleration of the CME and is also responsible for the expansion, we plotted the distribution of heights where the width became constant and the heights where the initial impulsive acceleration vanished (Figure \ref{exp}(b)). We found that both the distributions peaked at an overlapping height range of $2.5-3$ R$_{\odot}$. Thus, we believe that the effect of Lorentz force remained dominant in the initial phase of evolution till a height range of $2.5-3$ R$_{\odot}$. In this regard, we note that \citet{2010A&A...522A.100P} tracked a circular bubble shaped cavity of a CME using the GCS model, but with a further constraint that the legs of the model being co – incident (that is alpha being $0$, giving the ice cream cone shape). They studied the timing of bubble expansion and acceleration and its relation to the associated flare, and they do not talk about the height of influence of Lorentz force. In this work, we do not put any such constraints on the model parameters, and we further statistically find the height of influence of Lorentz force on the evolution of CMEs, by connecting width evolution and acceleration profile. To our knowledge this is the first time the observational impact of Lorentz force on CME kinematics is statistically shown by connecting true width evolution to the true acceleration profile using the GCS model.

\item We found latitudinal deflection suffered by a subset of the events we have studied (Figures \ref{defl}(a), (b), (c) and (d),). We found true deflection in $~31$ $\%$ (18 out of 59) cases ($~$9 $\%$ (5 out of 59) of those showed more than 10 degrees of deflection). A distribution of the latitudinal location of the source regions of these CMEs and of the PA of the CME leading edge converted to equivalent latitude showed that the most of the CMEs propagated non-radially and a majority of them apparently getting deflected towards the equator. But a distribution of the initial and final GCS latitudes did not show such strong signatures of equator-ward deflection. From the plot of final versus initial GCS latitude, we do find most of the CMEs showing deflections towards the equator, but the angle of deflection being small in most cases as mentioned earlier. Thus we conclude that the latitude versus PA equivalent latitude plot does not provide a conclusive evidence of deflection of CMEs, as the numbers largely suffer from projection effects, and a study with the true numbers (the true latitudes found from 3D reconstruction) give a better understanding. A plot of the change in latitude of the CMEs from the GCS model fitting (refer Figure \ref{defl}(b)), further showed that most of the CMEs got deflected towards the equator. The presence of coronal holes results in CMEs getting deflected and in our case too, coronal holes were found near the CMEs which were deflected by more than 10 degrees. \citet{Kay_2015} reported that such interaction of CMEs with the open field lines of coronal holes guides the CMEs towards the heliospheric current sheet. Such equator-ward deflection of fast CMEs can be potential indicators of strong Solar Energetic Particle (SEP) events happening at 1 AU \citep{Kahler_2012}, as such CME and coronal hole interactions indicate a possible interaction between the coronal hole magnetic field and the magnetic field of the CME, producing driverless shocks at 1 AU, capable of producing SEP events. Driverless shocks are the shocks at 1 AU which do not accompany any in-situ observation of the associated interplanetary CME drivers (for details, please refer \citet{Gopalswamy_2010}). Since we find only 5 cases which show strong equator-ward deflection, extending this study on a larger data set will help in confirming our conclusions.

\item We found the peak 3D speeds of CMEs ranging between 396 to 2465 km~s$^{-1}$ with a mean value of 984 km~s$^{-1}$, mode lying in the range 600 to 800 km~s$^{-1}$ and the peak 3D acceleration having an even wider range from 176 to 10922 m~s$^{-2}$ with a mean value of 2387 m~s$^{-2}$ and mode lying in the range 1000 to 2000 m~s$^{-2}$. We further found that the distribution of $V_{peak}$ and $a_{peak}$ showed a long tail towards high speeds and accelerations and this contribution to high values came from CMEs which had either ARs or APs as their source regions, thus showing that such CMEs are more energetic than the ones originating from PEs.  This conclusion is based on the speed and acceleration of these CMEs. A detailed investigation needs to be done considering the magnetic field of the source region and the injected flux to the CMEs to further quantify.

\item We studied the distribution of acceleration duration of CMEs and found a wide range from 10 minutes to 703 minutes with the peak of the distribution in the range of 100-150 minutes and the mean duration being 153 minutes. We also plotted the true acceleration magnitude versus acceleration duration and found a strong inverse correlation (correlation coefficient -0.88) between them with a power law index of -1.19, implying that stronger acceleration magnitudes have shorter acceleration duration. Further by comparing our result with previous results by \citet{Zhang_2006}, \citet{2011ApJ...738..191B} and \citet{2007SoPh..241...85V}, we found that this scaling relation between acceleration magnitude and acceleration duration remains similar in 3-D. Furthermore, we noted that the CMEs from PEs were mostly gradual events with weaker accelerations and longer acceleration duration. We also note that there would be acceleration below COR-1 FOV, and in such cases, the measurement of the acceleration phase would be an under estimate. But including EUV (to capture the acceleration below COR-1 FOV) will be tricky since we do not know if the structures seen in EUV map to the same features observed in white-light. Thus we confined the analysis to white light only. This can be done in future using data from Solar Orbiter \citep{2013SoPh..285...25M}, PROBA-3 \citep{proba-3} and ADITYA L1 \citep{aditya-l1,prasad_velc_2017}, which will observe the inner corona.
\item We compared the average values of true speeds and projected speeds in Figure \ref{vel_comp}. The average true speeds showed a wide range from 213 km~s$^{-1}$ to 1492 km~s$^{-1}$. We found that for almost all the slow CMEs, the 3D values are higher than the projected 2-D values. However, for the fast CMEs, a large fraction of the CMEs ($\sim~$63\%, (19 out of 30)) showed projected speeds greater than the actual 3D speeds. Thus we see a striking dissimilarity in the comparison of speeds for slow and fast CMEs. These fast CMEs were marked as partial-halo CMEs in the CDAW catalogue. A 3D reconstruction from multi-view points where the CME was non-halo gives the true picture with their true speeds. We also remind here that the method of tracking the CME height is different in our case and the one used in the LASCO images for the catalogue. One of the principal aim of studying CME kinematics is to calculate their arrival times at Earth, and it is clear from here that for the slow CMEs, the projected speeds will give a wrong estimation of the predicted arrival times. These considerations are important for better arrival time predictions with the true 3D speed for halo CMEs and not with the misleading projected speed which for halo CMEs might be the lateral expansion speed instead of the radial propagation speed.

\end{itemize}

We believe that this study will help us to understand and connect the true width evolution of CMEs in the inner corona, and the presence of impulsive acceleration in the 3D kinematic profiles will provide important inputs to CME ejection models. Comparison of projected and true speeds showed the importance of studying kinematics in 3D. We also point out that extending this work on a larger data set will help in confirming  our conclusions. It is also worth noting that the future solar missions, like ADITYA-L1 \citep{aditya-l1,prasad_velc_2017}, PROBA-3 \citep{proba-3} and the recently launched Solar Orbiter \citep{2013SoPh..285...25M} have coronagraphs that will be observing the inner corona. This work will provide essential inputs to plan the  observing campaigns of these missions and the data from these missions can be incorporated to extract more information on the kinematics of CMEs close to their initiation heights.


\acknowledgments
We thank the anonymous referee for valuable comments that have improved the manuscript. We would like to thank IIA for providing the computational facilities. VP was supported by the GOA-2015-014 (KU Leuven) and the European Research Council (ERC) under the European Union's Horizon 2020 research and innovation programme (grant agreement No 724326). We would like to thank Bibhuti Kumar Jha for his valuable suggestions.
The SECCHI data used here were produced by an international consortium of the Naval Research Laboratory (USA), Lockheed Martin Solar and Astrophysics Lab (USA), NASA Goddard Space Flight Center (USA), Rutherford Appleton Laboratory (UK), University of Birmingham (UK), Max-Planck-Institut for Solar System Research (Germany), Centre Spatiale de Li$\grave{e}$ge (Belgium), Institut d'Optique Th$\acute{e}$orique et Appliqu$\acute{e}$e (France), Institut d'Astrophysique Spatiale (France).
We also acknowledge SDO team to make AIA data available and SOHO team for EIT and LASCO data.

%

\vspace{5mm}

\bibliography{references}

\begin{thebibliography}{}
\expandafter\ifx\csname natexlab\endcsname\relax\def\natexlab#1{#1}\fi
\providecommand{\url}[1]{\href{#1}{#1}}

\bibitem[{{Balmaceda} {et~al.}(2018{\natexlab{a}}){Balmaceda}, {Vourlidas},
  {Stenborg}, \& {Dal Lago}}]{Balmaceda2018ApJ}
{Balmaceda}, L.~A., {Vourlidas}, A., {Stenborg}, G., \& {Dal Lago}, A.
  2018{\natexlab{a}}, \apj, 863, 57

\bibitem[{{Balmaceda} {et~al.}(2018{\natexlab{b}}){Balmaceda}, {Vourlidas},
  {Stenborg}, \& {Dal Lago}}]{2018ApJ...863...57B}
---. 2018{\natexlab{b}}, \apj, 863, 57

\bibitem[{{Bein} {et~al.}(2011){Bein}, {Berkebile-Stoiser}, {Veronig},
  {Temmer}, {Muhr}, {Kienreich}, {Utz}, \&
  {Vr{\v{s}}nak}}]{2011ApJ...738..191B}
{Bein}, B.~M., {Berkebile-Stoiser}, S., {Veronig}, A.~M., {et~al.} 2011, \apj,
  738, 191

\bibitem[{{Bein} {et~al.}(2012){Bein}, {Berkebile-Stoiser}, {Veronig},
  {Temmer}, \& {Vr{\v{s}}nak}}]{Bein2012ApJB}
{Bein}, B.~M., {Berkebile-Stoiser}, S., {Veronig}, A.~M., {Temmer}, M., \&
  {Vr{\v{s}}nak}, B. 2012, \apj, 755, 44

\bibitem[{{Bosman} {et~al.}(2012){Bosman}, {Bothmer}, {Nistic{\`o}},
  {Vourlidas}, {Howard}, \& {Davies}}]{2012SoPh..281..167B}
{Bosman}, E., {Bothmer}, V., {Nistic{\`o}}, G., {et~al.} 2012, \solphys, 281,
  167

\bibitem[{{Boursier} {et~al.}(2009){Boursier}, {Lamy}, {Llebaria}, {Goudail},
  \& {Robelus}}]{artemis}
{Boursier}, Y., {Lamy}, P., {Llebaria}, A., {Goudail}, F., \& {Robelus}, S.
  2009, \solphys, 257, 125

\bibitem[{{Brueckner} {et~al.}(1995){Brueckner}, {Howard}, {Koomen},
  {Korendyke}, {Michels}, {Moses}, {Socker}, {Dere}, {Lamy}, {Llebaria},
  {Bout}, {Schwenn}, {Simnett}, {Bedford}, \& {Eyles}}]{Brueckner95}
{Brueckner}, G.~E., {Howard}, R.~A., {Koomen}, M.~J., {et~al.} 1995, \solphys,
  162, 357

\bibitem[{{Burkepile} {et~al.}(2004){Burkepile}, {Hundhausen}, {Stanger}, {St.
  Cyr}, \& {Seiden}}]{Burkepile2004JGRAB}
{Burkepile}, J.~T., {Hundhausen}, A.~J., {Stanger}, A.~L., {St. Cyr}, O.~C., \&
  {Seiden}, J.~A. 2004, Journal of Geophysical Research (Space Physics), 109,
  A03103

\bibitem[{Byrne {et~al.}(2012)Byrne, Morgan, Habbal, \& Gallagher}]{corimp_2}
Byrne, J.~P., Morgan, H., Habbal, S.~R., \& Gallagher, P.~T. 2012, The
  Astrophysical Journal, 752, 145.
\newblock \url{https://doi.org/10.1088%2F0004-637x%2F752%2F2%2F145}

\bibitem[{{Cabello} {et~al.}(2016){Cabello}, {Cremades}, {Balmaceda}, \&
  {Dohmen}}]{Cabello2016SoPh}
{Cabello}, I., {Cremades}, H., {Balmaceda}, L., \& {Dohmen}, I. 2016, \solphys,
  291, 1799

\bibitem[{{Chen} {et~al.}(2006){Chen}, {Chen}, \& {Fang}}]{Chen2006A&AC}
{Chen}, A.~Q., {Chen}, P.~F., \& {Fang}, C. 2006, \aap, 456, 1153

\bibitem[{{Cheng} {et~al.}(2020){Cheng}, {Zhang}, {Kliem}, {\{T{\"o}r{\"o}k\}},
  {Xing}, {Zhou}, {Inhester}, \& {Ding}}]{2020arXiv200403790C}
{Cheng}, X., {Zhang}, J., {Kliem}, B., {et~al.} 2020, arXiv e-prints,
  arXiv:2004.03790

\bibitem[{{Cremades} \& {Bothmer}(2004)}]{Cremades2004A&A}
{Cremades}, H., \& {Bothmer}, V. 2004, \aap, 422, 307

\bibitem[{{Cremades} {et~al.}(2020){Cremades}, {Iglesias}, \&
  {Merenda}}]{2020A&A...635A.100C}
{Cremades}, H., {Iglesias}, F.~A., \& {Merenda}, L.~A. 2020, \aap, 635, A100

\bibitem[{{Delaboudini{\`e}re} {et~al.}(1995){Delaboudini{\`e}re}, {Artzner},
  {Brunaud}, {Gabriel}, {Hochedez}, {Millier}, {Song}, {Au}, {Dere}, {Howard},
  {Kreplin}, {Michels}, {Moses}, {Defise}, {Jamar}, {Rochus}, {Chauvineau},
  {Marioge}, {Catura}, {Lemen}, {Shing}, {Stern}, {Gurman}, {Neupert},
  {Maucherat}, {Clette}, {Cugnon}, \& {van Dessel}}]{SOHOEIT}
{Delaboudini{\`e}re}, J.-P., {Artzner}, G.~E., {Brunaud}, J., {et~al.} 1995,
  \solphys, 162, 291

\bibitem[{{Dere} {et~al.}(1997){Dere}, {Brueckner}, \&
  {Delaboudiniere}}]{Dere1997SPD}
{Dere}, K., {Brueckner}, G.~E., \& {Delaboudiniere}, J.~P. 1997, in AAS/Solar
  Physics Division Meeting \#28, AAS/Solar Physics Division Meeting, 05.02

\bibitem[{Gilbert {et~al.}(2000)Gilbert, Holzer, Burkepile, \&
  Hundhausen}]{Gilbert_2000}
Gilbert, H.~R., Holzer, T.~E., Burkepile, J.~T., \& Hundhausen, A.~J. 2000, The
  Astrophysical Journal, 537, 503.
\newblock \url{https://doi.org/10.1086%2F309030}

\bibitem[{{Gopalswamy}(2006)}]{Gopalswamy2006JApAG}
{Gopalswamy}, N. 2006, Journal of Astrophysics and Astronomy, 27, 243

\bibitem[{{Gopalswamy} {et~al.}(2009){Gopalswamy}, {Dal Lago}, {Yashiro}, \&
  {Akiyama}}]{2009CEAB...33..115G}
{Gopalswamy}, N., {Dal Lago}, A., {Yashiro}, S., \& {Akiyama}, S. 2009, Central
  European Astrophysical Bulletin, 33, 115

\bibitem[{Gopalswamy {et~al.}(2000)Gopalswamy, Lara, Lepping, Kaiser,
  Berdichevsky, \& St.~Cyr}]{nat2000}
Gopalswamy, N., Lara, A., Lepping, R., {et~al.} 2000, Geophysical Research
  Letters, 27, doi:10.1029/1999GL003639

\bibitem[{Gopalswamy {et~al.}(2009)Gopalswamy, Makela, Xie, Akiyama, \&
  Yashiro}]{nat}
Gopalswamy, N., Makela, P., Xie, H., Akiyama, S., \& Yashiro, S. 2009, Journal
  of Geophysical Research, 114, doi:10.1029/2008JA013686

\bibitem[{Gopalswamy {et~al.}(2003)Gopalswamy, Shimojo, Lu, Yashiro, Shibasaki,
  \& Howard}]{Gopalswamy_2003}
Gopalswamy, N., Shimojo, M., Lu, W., {et~al.} 2003, The Astrophysical Journal,
  586, 562.
\newblock \url{https://doi.org/10.1086%2F367614}

\bibitem[{Gopalswamy {et~al.}(2010)Gopalswamy, Xie, Mäkelä, Akiyama, Yashiro,
  Kaiser, Howard, \& Bougeret}]{Gopalswamy_2010}
Gopalswamy, N., Xie, H., Mäkelä, P., {et~al.} 2010, The Astrophysical
  Journal, 710, 1111.
\newblock \url{https://doi.org/10.1088%2F0004-637x%2F710%2F2%2F1111}

\bibitem[{{Gopalswamy} {et~al.}(2009){Gopalswamy}, {Yashiro}, {Michalek},
  {Stenborg}, {Vourlidas}, {Freeland}, \& {Howard}}]{Gopalswamy2009EM&PG}
{Gopalswamy}, N., {Yashiro}, S., {Michalek}, G., {et~al.} 2009, Earth Moon and
  Planets, 104, 295

\bibitem[{{Gosling}(1993)}]{1993JGR....9818937G}
{Gosling}, J.~T. 1993, \jgr, 98, 18937

\bibitem[{{Gui} {et~al.}(2011){Gui}, {Shen}, {Wang}, {Ye}, {Liu}, {Wang}, \&
  {Zhao}}]{2011SoPh..271..111G}
{Gui}, B., {Shen}, C., {Wang}, Y., {et~al.} 2011, \solphys, 271, 111

\bibitem[{{Hansen} {et~al.}(1971){Hansen}, {Garcia}, {Grognard}, \&
  {Sheridan}}]{1971PASAu...2...57H}
{Hansen}, R.~T., {Garcia}, C.~J., {Grognard}, R.~J.-M., \& {Sheridan}, K.~V.
  1971, Proceedings of the Astronomical Society of Australia, 2, 57

\bibitem[{Howard {et~al.}(2002)Howard, Moses, Socker, Dere, \& Cook}]{secchi}
Howard, R., Moses, J., Socker, D., Dere, K., \& Cook, J. 2002, Advances in
  Space Research, 29, 2017

\bibitem[{{Hundhausen} {et~al.}(1984){Hundhausen}, {Sawyer}, {House}, {Illing},
  \& {Wagner}}]{1984JGR....89.2639H}
{Hundhausen}, A.~J., {Sawyer}, C.~B., {House}, L., {Illing}, R.~M.~E., \&
  {Wagner}, W.~J. 1984, \jgr, 89, 2639

\bibitem[{{Hutton} \& {Morgan}(2017)}]{Hutton2017A&A}
{Hutton}, J., \& {Morgan}, H. 2017, \aap, 599, A68

\bibitem[{Isenberg \& Forbes(2007)}]{Isenberg_2007}
Isenberg, P.~A., \& Forbes, T.~G. 2007, The Astrophysical Journal, 670, 1453.
\newblock \url{https://doi.org/10.1086%2F522025}

\bibitem[{{Joshi} \& {Srivastava}(2011)}]{Joshi2011ApJ}
{Joshi}, A.~D., \& {Srivastava}, N. 2011, \apj, 739, 8

\bibitem[{Kahler {et~al.}(2012)Kahler, Akiyama, \& Gopalswamy}]{Kahler_2012}
Kahler, S.~W., Akiyama, S., \& Gopalswamy, N. 2012, The Astrophysical Journal,
  754, 100.
\newblock \url{https://doi.org/10.1088%2F0004-637x%2F754%2F2%2F100}

\bibitem[{{Kay} {et~al.}(2013){Kay}, {Opher}, \& {Evans}}]{2013ApJ...775....5K}
{Kay}, C., {Opher}, M., \& {Evans}, R.~M. 2013, \apj, 775, 5

\bibitem[{Kay {et~al.}(2015)Kay, Opher, \& Evans}]{Kay_2015}
Kay, C., Opher, M., \& Evans, R.~M. 2015, The Astrophysical Journal, 805, 168.
\newblock \url{https://doi.org/10.1088%2F0004-637x%2F805%2F2%2F168}

\bibitem[{Kliem {et~al.}(2014)Kliem, Lin, Forbes, Priest, \&
  Török}]{Kliem_2014}
Kliem, B., Lin, J., Forbes, T.~G., Priest, E.~R., \& Török, T. 2014, The
  Astrophysical Journal, 789, 46.
\newblock \url{https://doi.org/10.1088%2F0004-637x%2F789%2F1%2F46}

\bibitem[{Lemen {et~al.}(2011)Lemen, Title, Boerner, Chou, Drake, Duncan,
  Edwards, Friedlaender, Heyman, Hurlburt, Katz, Kushner, Levay, Lindgren,
  Mathur, McFeaters, Mitchell, Rehse, \& Waltham}]{aia}
Lemen, J., Title, A., Boerner, P., {et~al.} 2011, Solar Physics, 275, 17

\bibitem[{Lugaz {et~al.}(2012)Lugaz, Farrugia, Davies, Möstl, Davis, Roussev,
  \& Temmer}]{Lugaz_2012}
Lugaz, N., Farrugia, C.~J., Davies, J.~A., {et~al.} 2012, The Astrophysical
  Journal, 759, 68.
\newblock \url{https://doi.org/10.1088%2F0004-637x%2F759%2F1%2F68}

\bibitem[{{MacQueen} {et~al.}(1980){MacQueen}, {Csoeke-Poeckh}, {Hildner},
  {House}, {Reynolds}, {Stanger}, {Tepoel}, \& {Wagner}}]{SMM1980}
{MacQueen}, R.~M., {Csoeke-Poeckh}, A., {Hildner}, E., {et~al.} 1980, \solphys,
  65, 91

\bibitem[{{MacQueen} \& {Fisher}(1983)}]{MacQueen1983SoPhM}
{MacQueen}, R.~M., \& {Fisher}, R.~R. 1983, \solphys, 89, 89

\bibitem[{{Mierla} {et~al.}(2009){Mierla}, {Inhester}, {Marqu{\'e}},
  {Rodriguez}, {Gissot}, {Zhukov}, {Berghmans}, \& {Davila}}]{Mierla2009SoPh}
{Mierla}, M., {Inhester}, B., {Marqu{\'e}}, C., {et~al.} 2009, \solphys, 259,
  123

\bibitem[{{Mierla} {et~al.}(2011){Mierla}, {Inhester}, {Rodriguez}, {Gissot},
  {Zhukov}, \& {Srivastava}}]{Mierla2011JASTP}
{Mierla}, M., {Inhester}, B., {Rodriguez}, L., {et~al.} 2011, Journal of
  Atmospheric and Solar-Terrestrial Physics, 73, 1166

\bibitem[{{Mierla} {et~al.}(2008){Mierla}, {Davila}, {Thompson}, {Inhester},
  {Srivastava}, {Kramar}, {St. Cyr}, {Stenborg}, \& {Howard}}]{Mierla2008SoPh}
{Mierla}, M., {Davila}, J., {Thompson}, W., {et~al.} 2008, \solphys, 252, 385

\bibitem[{{Mierla} {et~al.}(2010){Mierla}, {Inhester}, {Antunes}, {Boursier},
  {Byrne}, {Colaninno}, {Davila}, {de Koning}, {Gallagher}, {Gissot}, {Howard},
  {Howard}, {Kramar}, {Lamy}, {Liewer}, {Maloney}, {Marqu{\'e}}, {McAteer},
  {Moran}, {Rodriguez}, {Srivastava}, {St. Cyr}, {Stenborg}, {Temmer},
  {Thernisien}, {Vourlidas}, {West}, {Wood}, \& {Zhukov}}]{Mierla2010AnGeo}
{Mierla}, M., {Inhester}, B., {Antunes}, A., {et~al.} 2010, Annales
  Geophysicae, 28, 203

\bibitem[{{Mierla} {et~al.}(2013){Mierla}, {Seaton}, {Berghmans}, {Chifu}, {De
  Groof}, {Inhester}, {Rodriguez}, {Stenborg}, \&
  {Zhukov}}]{2013SoPh..286..241M}
{Mierla}, M., {Seaton}, D.~B., {Berghmans}, D., {et~al.} 2013, \solphys, 286,
  241

\bibitem[{{Moon} {et~al.}(2002){Moon}, {Choe}, {Wang}, {Park}, {Gopalswamy},
  {Yang}, \& {Yashiro}}]{Moon2002ApJM}
{Moon}, Y.~J., {Choe}, G.~S., {Wang}, H., {et~al.} 2002, \apj, 581, 694

\bibitem[{{Moran} {et~al.}(2010){Moran}, {Davila}, \&
  {Thompson}}]{Moran2010ApJ}
{Moran}, T.~G., {Davila}, J.~M., \& {Thompson}, W.~T. 2010, \apj, 712, 453

\bibitem[{Morgan {et~al.}(2012)Morgan, Byrne, \& Habbal}]{corimp_1}
Morgan, H., Byrne, J.~P., \& Habbal, S.~R. 2012, The Astrophysical Journal,
  752, 144.
\newblock \url{https://doi.org/10.1088%2F0004-637x%2F752%2F2%2F144}

\bibitem[{{M{\"u}ller} {et~al.}(2013){M{\"u}ller}, {Marsden}, {St. Cyr}, \&
  {Gilbert}}]{2013SoPh..285...25M}
{M{\"u}ller}, D., {Marsden}, R.~G., {St. Cyr}, O.~C., \& {Gilbert}, H.~R. 2013,
  \solphys, 285, 25

\bibitem[{{Muller} {et~al.}(2009){Muller}, {Fleck}, {Dimitoglou}, {Caplins},
  {Amadigwe}, {Garc{\'\i}a Ortiz}, {Wamsler}, {Alexanderian}, {Hughitt}, \&
  {Ireland}}]{jhelio2}
{Muller}, D., {Fleck}, B., {Dimitoglou}, G., {et~al.} 2009, Computing in
  Science and Engineering, 11, 38

\bibitem[{{M{\"u}ller} {et~al.}(2017){M{\"u}ller}, {Nicula}, {Felix},
  {Verstringe}, {Bourgoignie}, {Csillaghy}, {Berghmans}, {Jiggens},
  {Garc{\'\i}a-Ortiz}, {Ireland}, {Zahniy}, \& {Fleck}}]{jhelio}
{M{\"u}ller}, D., {Nicula}, B., {Felix}, S., {et~al.} 2017, \aap, 606, A10

\bibitem[{{Murray} {et~al.}(2018){Murray}, {Guerra}, {Zucca}, {Park}, {Carley},
  {Gallagher}, {Vilmer}, \& {Bothmer}}]{2018SoPh..293...60M}
{Murray}, S.~A., {Guerra}, J.~A., {Zucca}, P., {et~al.} 2018, \solphys, 293, 60

\bibitem[{{Olmedo} {et~al.}(2008){Olmedo}, {Zhang}, {Wechsler}, {Poland}, \&
  {Borne}}]{seeds}
{Olmedo}, O., {Zhang}, J., {Wechsler}, H., {Poland}, A., \& {Borne}, K. 2008,
  \solphys, 248, 485

\bibitem[{{Pant} {et~al.}(2016){Pant}, {Willems}, {Rodriguez}, {Mierla},
  {Banerjee}, \& {Davies}}]{2016ApJ...833...80P}
{Pant}, V., {Willems}, S., {Rodriguez}, L., {et~al.} 2016, \apj, 833, 80

\bibitem[{{Patsourakos} {et~al.}(2010){Patsourakos}, {Vourlidas}, \&
  {Kliem}}]{2010A&A...522A.100P}
{Patsourakos}, S., {Vourlidas}, A., \& {Kliem}, B. 2010, \aap, 522, A100

\bibitem[{Prasad {et~al.}(2017)Prasad, Banerjee, Singh, Subramanya, Kumar,
  Kamath, Kathiravan, Venkata, Rajkumar, Natarajan, Juneja, Somu, Pant, Shaji,
  Sankarsubramanian, Patra, Venkateswaran, Adoni, Narendra, \&
  Jaiswal}]{prasad_velc_2017}
Prasad, B., Banerjee, D., Singh, J., {et~al.} 2017, Current Science, 113, 613

\bibitem[{{R Core Team}(2018)}]{R}
{R Core Team}. 2018, R: A Language and Environment for Statistical Computing, R
  Foundation for Statistical Computing, Vienna, Austria.
\newblock \url{https://www.R-project.org/}

\bibitem[{Renotte {et~al.}(2014)Renotte, Baston, Bemporad, Capobianco, Cernica,
  Darakchiev, Denis, Desselle, Vos, Fineschi, Focardi, Górski, Graczyk,
  Halain, Hermans, Jackson, Kintziger, Kosiec, Kranitis, \& Zhukov}]{proba-3}
Renotte, E., Baston, E., Bemporad, A., {et~al.} 2014, in , 91432M

\bibitem[{{Robbrecht} \& {Berghmans}(2004)}]{cactus}
{Robbrecht}, E., \& {Berghmans}, D. 2004, \aap, 425, 1097

\bibitem[{Sachdeva {et~al.}(2015)Sachdeva, Subramanian, Colaninno, \&
  Vourlidas}]{Sachdeva_2015}
Sachdeva, N., Subramanian, P., Colaninno, R., \& Vourlidas, A. 2015, The
  Astrophysical Journal, 809, 158.
\newblock \url{https://doi.org/10.1088%2F0004-637x%2F809%2F2%2F158}

\bibitem[{{Sachdeva} {et~al.}(2017){Sachdeva}, {Subramanian}, {Vourlidas}, \&
  {Bothmer}}]{2017SoPh..292..118S}
{Sachdeva}, N., {Subramanian}, P., {Vourlidas}, A., \& {Bothmer}, V. 2017,
  \solphys, 292, 118

\bibitem[{{Sarkar} {et~al.}(2019){Sarkar}, {Srivastava}, {Mierla}, {West}, \&
  {D'Huys}}]{Sarkar2019ApJ}
{Sarkar}, R., {Srivastava}, N., {Mierla}, M., {West}, M.~J., \& {D'Huys}, E.
  2019, \apj, 875, 101

\bibitem[{{Schwenn}(2006)}]{2006LRSP....3....2S}
{Schwenn}, R. 2006, Living Reviews in Solar Physics, 3, 2

\bibitem[{Seetha \& Megala(2017)}]{aditya-l1}
Seetha, S., \& Megala, S. 2017, Current Science, 113, 610

\bibitem[{{Sheeley} {et~al.}(1999){Sheeley}, {Walters}, {Wang}, \&
  {Howard}}]{Sheeley1999JGR}
{Sheeley}, N.~R., {Walters}, J.~H., {Wang}, Y.~M., \& {Howard}, R.~A. 1999,
  \jgr, 104, 24739

\bibitem[{{Shen} {et~al.}(2013{\natexlab{a}}){Shen}, {Wang}, {Pan}, {Zhang},
  {Ye}, \& {Wang}}]{2013AGUFMSH31B2034S}
{Shen}, C., {Wang}, Y., {Pan}, Z., {et~al.} 2013{\natexlab{a}}, in AGU Fall
  Meeting Abstracts, Vol. 2013, SH31B--2034

\bibitem[{{Shen} {et~al.}(2013{\natexlab{b}}){Shen}, {Wang}, {Pan}, {Zhang},
  {Ye}, \& {Wang}}]{2013JGRA..118.6858S}
{Shen}, C., {Wang}, Y., {Pan}, Z., {et~al.} 2013{\natexlab{b}}, Journal of
  Geophysical Research (Space Physics), 118, 6858

\bibitem[{{St. Cyr} {et~al.}(1999){St. Cyr}, {Burkepile}, {Hundhausen}, \&
  {Lecinski}}]{Cyr1999JGR}
{St. Cyr}, O.~C., {Burkepile}, J.~T., {Hundhausen}, A.~J., \& {Lecinski}, A.~R.
  1999, \jgr, 104, 12493

\bibitem[{{St. Cyr} {et~al.}(2000){St. Cyr}, {Plunkett}, {Michels},
  {Paswaters}, {Koomen}, {Simnett}, {Thompson}, {Gurman}, {Schwenn}, {Webb},
  {Hildner}, \& {Lamy}}]{StCyr2000JGRS}
{St. Cyr}, O.~C., {Plunkett}, S.~P., {Michels}, D.~J., {et~al.} 2000, \jgr,
  105, 18169

\bibitem[{Subramanian {et~al.}(2014)Subramanian, Arunbabu, Vourlidas, \&
  Mauriya}]{Subramanian_2014}
Subramanian, P., Arunbabu, K.~P., Vourlidas, A., \& Mauriya, A. 2014, The
  Astrophysical Journal, 790, 125.
\newblock \url{https://doi.org/10.1088%2F0004-637x%2F790%2F2%2F125}

\bibitem[{{Subramanian} \& {Dere}(2001)}]{2001ApJ...561..372S}
{Subramanian}, P., \& {Dere}, K.~P. 2001, \apj, 561, 372

\bibitem[{{Suryanarayana}(2019)}]{Suryanarayana2019JASTP}
{Suryanarayana}, G.~S. 2019, Journal of Atmospheric and Solar-Terrestrial
  Physics, 185, 1

\bibitem[{{Temmer}(2016)}]{Temmer2016AN}
{Temmer}, M. 2016, Astronomische Nachrichten, 337, 1010

\bibitem[{Temmer {et~al.}(2010)Temmer, Veronig, Kontar, Krucker, \&
  Vr{\v{s}}nak}]{Temmer_2010}
Temmer, M., Veronig, A.~M., Kontar, E.~P., Krucker, S., \& Vr{\v{s}}nak, B.
  2010, The Astrophysical Journal, 712, 1410.
\newblock \url{https://doi.org/10.1088%2F0004-637x%2F712%2F2%2F1410}

\bibitem[{{Thernisien}(2011)}]{Thernisien2011ApJST}
{Thernisien}, A. 2011, \apjs, 194, 33

\bibitem[{{Thernisien} {et~al.}(2009){Thernisien}, {Vourlidas}, \&
  {Howard}}]{2009SoPh..256..111T}
{Thernisien}, A., {Vourlidas}, A., \& {Howard}, R.~A. 2009, \solphys, 256, 111

\bibitem[{{Thernisien} {et~al.}(2006){Thernisien}, {Howard}, \&
  {Vourlidas}}]{Thernisien2006ApJ}
{Thernisien}, A.~F.~R., {Howard}, R.~A., \& {Vourlidas}, A. 2006, \apj, 652,
  763

\bibitem[{Vourlidas {et~al.}(2017)Vourlidas, Balmaceda, Stenborg, \&
  Lago}]{jhuapl}
Vourlidas, A., Balmaceda, L.~A., Stenborg, G., \& Lago, A.~D. 2017, The
  Astrophysical Journal, 838, 141.
\newblock \url{https://doi.org/10.3847%2F1538-4357%2Faa67f0}

\bibitem[{{Vr{\v{s}}nak} {et~al.}(2007){Vr{\v{s}}nak}, {Mari{\v{c}}i{\'c}},
  {Stanger}, {Veronig}, {Temmer}, \& {Ro{\v{s}}a}}]{2007SoPh..241...85V}
{Vr{\v{s}}nak}, B., {Mari{\v{c}}i{\'c}}, D., {Stanger}, A.~L., {et~al.} 2007,
  \solphys, 241, 85

\bibitem[{{Vr{\v{s}}nak} {et~al.}(2004){Vr{\v{s}}nak}, {Ru{\v{z}}djak},
  {Sudar}, \& {Gopalswamy}}]{vrsnak2004A&A}
{Vr{\v{s}}nak}, B., {Ru{\v{z}}djak}, D., {Sudar}, D., \& {Gopalswamy}, N. 2004,
  \aap, 423, 717

\bibitem[{{Vr{\v{s}}nak} {et~al.}(2005){Vr{\v{s}}nak}, {Sudar}, \&
  {Ru{\v{z}}djak}}]{Vrsnak2005A&AV}
{Vr{\v{s}}nak}, B., {Sudar}, D., \& {Ru{\v{z}}djak}, D. 2005, \aap, 435, 1149

\bibitem[{{Wang} {et~al.}(2019){Wang}, {Hoeksema}, \&
  {Liu}}]{2019arXiv190906410W}
{Wang}, J., {Hoeksema}, J.~T., \& {Liu}, S. 2019, arXiv e-prints,
  arXiv:1909.06410

\bibitem[{Webb \& Howard(2012)}]{article}
Webb, D., \& Howard, T. 2012, Living Reviews in Solar Physics, 9, 3

\bibitem[{{Wood} {et~al.}(1999){Wood}, {Karovska}, {Chen}, {Brueckner}, {Cook},
  \& {Howard}}]{Wood1999ApJW}
{Wood}, B.~E., {Karovska}, M., {Chen}, J., {et~al.} 1999, \apj, 512, 484

\bibitem[{{Yashiro} {et~al.}(2004){Yashiro}, {Gopalswamy}, {Michalek},
  {St.~Cyr}, {Plunkett}, {Rich}, \& {Howard}}]{Yashiro04}
{Yashiro}, S., {Gopalswamy}, N., {Michalek}, G., {et~al.} 2004, Journal of
  Geophysical Research (Space Physics), 109, A07105

\bibitem[{Zhang \& Dere(2006)}]{Zhang_2006}
Zhang, J., \& Dere, K.~P. 2006, The Astrophysical Journal, 649, 1100.
\newblock \url{https://doi.org/10.1086%2F506903}

\bibitem[{{Zhang} {et~al.}(2001){Zhang}, {Dere}, {Howard}, {Kundu}, \&
  {White}}]{Zhang2001}
{Zhang}, J., {Dere}, K.~P., {Howard}, R.~A., {Kundu}, M.~R., \& {White}, S.~M.
  2001, \apj, 559, 452

\bibitem[{Zhang {et~al.}(2004)Zhang, Dere, Howard, \& Vourlidas}]{Zhang_2004}
Zhang, J., Dere, K.~P., Howard, R.~A., \& Vourlidas, A. 2004, The Astrophysical
  Journal, 604, 420.
\newblock \url{https://doi.org/10.1086%2F381725}

\end{thebibliography}



\end{document}